\documentclass[onecolumn,useAMS,usenatbib]{mn2e}

\usepackage{graphicx,amsmath}

\usepackage{natbib}

\usepackage{bm}

\title[GGI self-calibration]
{Self-Calibration Technique for 3-point Intrinsic Alignment Correlations in Weak Lensing Surveys} 

\author[Troxel \& Ishak]
 {M. A. Troxel\thanks{Electronic address:
    {\tt troxel@utdallas.edu}} and
  M. Ishak\thanks{Electronic address:
    {\tt mishak@utdallas.edu}}$^1$ 
\\$^1$Department of Physics, The University of Texas at Dallas, Richardson, TX 75083, USA
}

\begin{document}

\date{\today}

\pagerange{\pageref{firstpage}--\pageref{lastpage}} \pubyear{0000}

\maketitle

\label{firstpage}

\begin{abstract}
The intrinsic alignment (IA) of galaxies has been shown to be a significant barrier to precision cosmic shear measurements. (Zhang, 2010, ApJ, 720, 1090) proposed a self-calibration technique for the power spectrum to calculate the induced gravitational shear-galaxy intrinsic ellipticity correlation (GI) in weak lensing surveys with photo-z measurements which is expected to reduce the IA contamination by at least a factor of 10 for currently proposed surveys. We confirm this using an independent analysis and propose an expansion to the self-calibration technique for the bispectrum in order to calculate the dominant IA gravitational shear-gravitational shear-intrinsic ellipticity correlation (GGI) contamination. We first establish an estimator to extract the galaxy density-density-intrinsic ellipticity (ggI) correlation from the galaxy ellipticity-density-density measurement for a photo-z galaxy sample. We then develop a relation between the GGI and ggI bispectra, which allows for the estimation and removal of the GGI correlation from the cosmic shear signal. We explore the performance of these two methods, compare to other possible sources of error, and show that the GGI self-calibration technique can potentially reduce the IA contamination by up to a factor of 5-10 for all but a few bin choices, thus reducing the contamination to the percent level. The self-calibration is less accurate for adjacent bins, but still allows for a factor of three reduction in the IA contamination. The self-calibration thus promises to be an efficient technique to isolate both the 2-point and 3-point intrinsic alignment signals from weak lensing measurements.

\end{abstract}


\section{Introduction}\label{intro}

Weak gravitational lensing due to large scale structure (cosmic shear) has emerged as a powerful cosmological probe in order to map the distribution of dark matter in the universe and to characterise the equation of state of dark energy, improving constraints on the equation of state of dark energy and the matter fluctuation amplitude parameter by factors of 2 to 4 (see for example \cite{1a,1b,1c,1d,1e,1f,1g,1h,1i,1j,1k,1l,1m,1n,1o,1p} and references therein.) Gravitational lensing has also been shown to be very useful to test the nature of gravity at cosmological distance scales (see for example the partial list \cite{2a,2b,2c,2d,2e,2f,2g,2h,2i,2j,2k,2l,2m,2n,2o,2p,2q}.)

In addition to the constraints obtained from the 2-point cosmic shear correlation and the corresponding shear power spectrum, the 3-point cosmic shear correlation and the shear bispectrum have been shown to break degeneracies in the cosmological parameters that the power spectrum alone does not \citep{3a,3b}. For example, the results of \cite{4} showed that a deep lensing survey should be able to improve the constraints on the dark energy parameters and the matter fluctuation amplitude by a further factor of 2-3 using the bispectrum, and most recently, \cite{5} derived parameter constraints by measuring the third order moment of the aperture mass measure using weak lensing data from the HST COSMOS survey. They found independent results consistent with WMAP7 best-fit cosmology, and an improved constraint when combined with the 2-point correlation. Ongoing, future and proposed lensing surveys (e.g. CFHTLS\footnote[1]{\slshape http://www.cfht.hawaii.edu/Science/CFHLS/}, DES\footnote[2]{\slshape http://www.darkenergysurvey.org/}, EUCLID\footnote[3]{\slshape http://sci.esa.int/euclid/}, HSC\footnote[4]{\slshape http://www.naoj.org/Projects/HSC/}, HST\footnote[5]{\slshape http://www.stsci.edu/hst/}, JWST\footnote[6]{\slshape http://www.jwst.nasa.gov/}, LSST\footnote[7]{\slshape http://www.lsst.org/lsst/}, Pan-STARRS\footnote[8]{\slshape http://pan-starrs.ifa.hawaii.edu/}, and WFIRST\footnote[9]{\slshape http://wfirst.gsfc.nasa.gov/}) promise to provide precision cosmic shear measurements.

Cosmic shear measurements are limited in precision by several systematic effects. It is important to understand and control these systematic effects in order to fully explore the potential of this probe (see for example \cite{6a,6b,6c,6d,6e,6f,6g,6h,6i,6j,6k,6l,6m,7} and references therein). One of the serious systematic effects of lensing is the correlated intrinsic alignment of galaxies which contaminates the lensing signal and acts as a nuisance factor (see for example \cite{7,8a,6c,8c,6d,8e,8f,8g,8h,9a,9b,10,11,12,13,14a,14b,15,16,17,6j,18b,19} and references therein). For example, \cite{9a,9b} showed that if intrinsic alignment is ignored the determination of the dark energy equation of state is biased by as much as $50\%$. \cite{10} found that the matter power spectrum amplitude can be affected by intrinsic alignment by up to $30\%$, showing the importance of developing methods to isolate the intrinsic alignment and remove it from the cosmic shear signal.

There are two 2-point intrinsic alignment correlations. The first is a correlation between the intrinsic ellipticity of two galaxies, known as the II correlation. If the two galaxies are spatially close, they can be aligned by the tidal force field of the same nearby matter structure. The second intrinsic alignment correlation, known as the GI correlation, was identified by \cite{11} and is due to a matter structure both causing the alignment of a nearby galaxy and contributing to the lensing signal of a background galaxy. This produces an anti-correlation between the cosmic shear and intrinsic ellipticity, since the tidal force and gravitational lensing tend to align the galaxy shapes in orthogonal directions. The GI correlation has been measured in various subsets of the SDSS spectroscopic and imaging samples by various groups. A detection of the large-scale GI correlation in the SDSS was reported by \cite{12} and then \cite{10} found an even stronger GI correlation for Luminous Red Galaxies (LRGs). It was shown in these papers that this contamination can affect the lensing measurement and cosmology up to the $10\%$ level and up to $30\%$ in some cases for the matter fluctuation amplitude. This finding was confirmed by numerical simulations, where a level of contamination of $10\%$ was found \citep{13}. Further measurements of the GI correlation were made in the SDSS dataset by \cite{14a,14b}. Most recently, \cite{15} measured strong 2-point intrinsic alignment correlations in various SDSS and MegaZ-LRG samples.

In a similar way, when we consider three galaxies and the related 3-point correlation, the cosmic shear signal (GGG bispectrum) also suffers from contamination by the 3-point intrinsic alignment correlations. The first is the III correlation between intrinsic ellipticities of three spatially close galaxies which are intrinsically aligned by a nearby matter structure. The second is the GII correlation, where two spatially close galaxies are intrinsically aligned by a nearby matter structure which contributes to the lensing of a third galaxy in the background. Finally, there is the GGI correlation, where two galaxies are lensed by a structure which intrinsically aligns a third galaxy in the foreground. Unlike the 2-point correlations, the sign of the GGI and GII correlations depend both on triangle shape and scale. \cite{16} showed that lensing bispectrum measurements are typically more strongly contaminated by intrinsic alignment compared to the lensing spectrum measurements, and that the contamination from the 3-point intrinsic alignment correlation can be as large as $15-20\%$ compared to the GGG lensing signal. Finally, 3-point intrinsic alignment measurements are not only useful for constraining their contamination to 3-point lensing measurements, but are also useful for constraining models of intrinsic alignments and therefore constraining the contamination to all lensing measurements (including 2-point correlations) which will dominate the science cases of upcoming surveys.

While the II and III intrinsic alignment correlations can be greatly reduced with photo-z's by using cross-spectra of galaxies in two different redshift bins (see for example \cite{6k}) so that the galaxies are separated by large enough distances to assure that the tidal effect is weak, this does not work for the GI, GGI, and GII correlations which happen between galaxies at different redshifts and large separations. The GI correlation and methods for its removal have been the topic of several recent scientific publications and we review these briefly. Initially, some first suggestions were discussed by \cite{11}. \cite{17} extended the approach of template fitting by \cite{6j,18b} to include a treatment of the GI correlation. \cite{9a,9b} investigated the effects of the GI correlation on cosmological parameter constraints by assuming a model of the GI intrinsic alignment that is binned in redshift and angular frequency with some free parameters that are marginalised over. \cite{19} performed a cosmological constraint analysis where modelling of intrinsic alignment was included, showing a significant effect on the amplitude of matter fluctuations. Using a geometrical approach, \cite{20a,20b,20c} proposed a nulling technique to remove the GI intrinsic alignment contribution by exploiting the redshift dependence of the correlations, but it is found that the technique throws out some of the valuable lensing signal. Most recently, the nulling technique has been applied at the 3-point level for the GGI correlation, but again with similar signal loss to that at the 2-point level \citep{21}.

Finally, \cite{22} proposed a technique to self-calibrate the GI intrinsic alignment signal by using the intrinsic galaxy ellipticity-galaxy density correlation, which requires that in addition to the galaxy ellipticity-ellipticity correlation (cosmic shear), one should also extract galaxy density-density and galaxy ellipticity-density correlations from the lensing survey. The GI correlation is then calculated and removed from the lensing signal. Most recently, \cite{23} showed that redshift dependencies of intrinsic alignment can allow further improvements to the calculation of the intrinsic alignment contamination. The technique is commonly referred to as self-calibration because it uses correlations that can be extracted from the same gravitational lensing survey and used in order to calculate the GI contamination to the cosmic shear signal and remove it. \cite{9b} applied an approach like the self-calibration, using correlations between lensing, intrinsic alignment, number density and magnification effects to constrain cosmological parameters. They found that the extra information from the additional correlations can make up for the additional free parameters in the intrinsic alignment so that the contamination can be removed without loss of constraining power.

We organize the paper as follows. In Sec. \ref{back}, we briefly discuss the necessary survey parameters and lensing formalism. We then provide a summary of the 2-point GI self-calibration technique of Zhang with independent results. In Sec. \ref{sc}, we develop the 3-point GGI self-calibration. We first establish an estimator to extract the galaxy density-density-intrinsic ellipticity correlation (ggI) from the observed galaxy ellipticity-density-density measurement for a photo-z galaxy sample. We then develop a relation between the GGI and ggI bispectra, which allows for the estimation and removal of the GGI intrinsic alignment correlation from the cosmic shear signal. Section \ref{error} describes the residual sources of error to the GGI self-calibration technique, and we present the necessary relations to quantify these errors. These are compared to other sources of error in the bispectrum. Finally, we summarise the effectiveness and impact of the GGI self-calibration in Sec. \ref{conc}. In the Appendix we expand upon the detailed calculation of the coefficients in the error calculation found in Sec. \ref{ciggerror} and provide a list of typical expected values.

%%%%%%%%%%%%%%%%%%%
%%%%%%%%%%%%%%%%%%%%%%%%%%%%%%%%%%%%%%%%%%%%%%%%%%%%
\section{Background}\label{back}
%%%%%%%%%%%%%%%%%%%%%%%%%%%%%%%%%%%%%%%%%%%%%%%%%%%%%
%%%%%%%%%%%%%%%%%%%
\subsection{Survey information and weak lensing}\label{survey}
%%%%%%%%%%%%%%%%%%%
As mentioned in the previous section, the self-calibration technique proposed by \cite{22} makes use of the information already found in a lensing survey \citep{25}, including galaxy shape, angular position and photometric redshift, in order to calculate and remove the dominant intrinsic alignment contamination. In our performance calculations, we consider survey parameters to match a survey similar to those of the LSST lensing survey \citep{29}, but this is just an example. The GGI self-calibration technique is survey independent and can be applied to reduce the intrinsic alignment contamination in all lensing surveys. Galaxies are assumed to be sufficiently large and bright to be suitable for cosmic shear measurements, so we restrict any discussion of the self-calibration to these galaxies in order to avoid any sample bias. Galaxies are split into photo-z bins according to photo-z $z^P$, where the \emph{i}-th photo-z bin is described by a mean photo-z $\bar{z}_i$ and has a range $\bar{z}_i-\Delta z_i/2\le z^P\le \bar{z}_i+\Delta z_i/2$. In our notation, $i<j$ implies that $\bar{z}_i<\bar{z}_j$. The galaxy redshift distribution over the \emph{i}-th redshift bin is $n^P_i(z^P)$ and $n_i(z)$ as a function of photo-z and true redshift, respectively, which are related by the photo-z probability distribution function $p(z|z^P)$. 

In evaluating the performance of the self-calibration technique, we will consider as an example survey parameters to match an LSST-like weak lensing survey \citep{29}, but of course the calculations are applicable to all current and planned weak lensing surveys (e.g. CFHTLS, DES, EUCLID, HSC, HST, JWST, LSST, Pan-STARRS, and WFIRST). We assume a survey coverage of half the sky ($f_{sky}=0.5$) with a total galaxy surface density of 40 arcminute$^{-2}$ and redshift density distribution of

\begin{equation}
n(z)=\frac{1}{2z_0}\left(\frac{z}{z_0}\right)^2\exp(-z/z_0),
\end{equation}
with $z_0=0.5$. The ellipticity shape noise is described by $\gamma_{rms}=0.18+0.042z$ and photo-z error by a Gaussian probability distribution function (PDF) 

\begin{equation}
p(z|z^P)=\frac{1}{\sqrt{2\pi\sigma_z^2}}\exp\left(\frac{-(z-z^P)^2}{2\sigma_z^2}\right),
\end{equation}
with $\sigma_z=0.05(1+z)$. We define photometric redshift bins of width $\Delta z=0.2$, centred at $\bar{z}_i=0.2(i+1)$ ($i=1,\cdots,9$). We do not include redshifts below $z^P=0.3$, not because of poor performance in the self-calibration, but rather due to the weaker lensing signal at lower redshifts. This artificially increases the fractional errors we evaluate in Sec. \ref{error} with respect to the GGG lensing signal, as is evident in the increasing errors at low redshift in Fig. \ref{fig:merror}, and is not useful in evaluating the true performance of the self-calibration.

The self-calibration technique relies upon two basic observables from a weak lensing survey. The first is the galaxy surface density of a photo-z bin, $\delta^{\Sigma}$, which is a function of the 3D galaxy distribution $\delta_g$. The second necessary observable is the galaxy shape, expressed in terms of ellipticity, which measures the cosmic shear $\gamma$. However, this cosmic shear signal is heavily contaminated by the intrinsic ellipticities of galaxies. There is a random component to this intrinsic ellipticity, which is simple to correct and which we ignore in the self-calibration calculations except as part of the shot noise in the error estimations of Sec. \ref{error}. A second component of the intrinsic ellipticity is due to the intrinsic alignment of galaxies caused by the gravitational tidal forces of large scale structure and was introduced in Sec. \ref{intro}.

We will label the measured shear as $\gamma^s=\gamma+\gamma^I$, where $\gamma^I$ denotes the correlated part of this intrinsic ellipticity due to the intrinsic alignment of galaxies. Since we are concerned only with the weak limit, we will work with the lensing convergence $\kappa$ instead. Thus from the measured $\gamma^s$, we obtain $\kappa^s=\kappa+\kappa^I$. $\kappa$ is the projected matter over-density along the line of sight. For a flat universe in the Born approximation, the convergence $\kappa$ of a source galaxy at redshift $z_G$ and direction $\hat{\theta}$ is

\begin{equation}
\kappa(\hat{\theta})=\int_0^{\chi_G}\delta(\chi_L,\hat{\theta})W_L(\chi_L,\chi_G)d\chi_L.
\end{equation}
$W_L(\chi_L,\chi_G)$ is the lensing kernel and $\delta(\chi_L,\hat{\theta})$ is the matter over-density in direction $\hat{\theta}$ and at comoving distance $\chi_L\equiv \chi_L(z_L)$. $\chi_G\equiv \chi_G(z_G)$ is the comoving distance to the source. The comoving distance $\chi$ is in units of $c/H_0$, where $H_0$ is the current day Hubble constant. We assume a flat $\Lambda$CDM cosmology with $\Omega_m=0.27$ and $\Omega_{\Lambda}=0.73$. The lensing kernel is then

\begin{equation}
W_L(\chi_L,\chi_G)=\frac{3}{2}\Omega_m(1+z_L)\chi_L(1-\frac{\chi_L}{\chi_G})
\end{equation}
when $z_L<z_G$ and zero otherwise. 

In our calculations, we will work in Fourier (multipole $\ell$) space with the corresponding spectra to the correlations which can be built from these survey observables. The two-point correlation function is then related to the angular power spectrum and the 3-point correlation function to the angular bispectrum by

\begin{eqnarray}
\langle \tilde{\kappa}(\bm{\ell_1})\tilde{\kappa}(\bm{\ell_2})\rangle&=&(2\pi)^2\delta^{D}(\bm{\ell_1}+\bm{\ell_2})C(\ell_1)\nonumber\\
\langle \tilde{\kappa}(\bm{\ell_1})\tilde{\kappa}(\bm{\ell_2})\tilde{\kappa}(\bm{\ell_3})\rangle&=&(2\pi)^2\delta^{D}(\bm{\ell_1}+\bm{\ell_2}+\bm{\ell_3})B(\ell_1,\ell_2,\ell_3),
\end{eqnarray}
where $\langle\cdots\rangle$ denotes the ensemble average and $\delta^{D}(\bm{\ell})$ is the Dirac delta function. For the bispectrum, $\delta^{D}(\bm{\ell}_1+\bm{\ell}_2+\bm{\ell}_3)$ enforces the condition that the three vectors form a triangle in Fourier space. The 2D angular cross-correlation power spectrum is related to the 3D power spectrum and the 2D angular cross-correlation bispectrum to the 3D bispectrum through the Limber approximation

\begin{eqnarray}
C_{ij}(\ell)&=&\int_0^{\chi}\frac{W_{i}(\chi')W_{j}(\chi')}{\chi'^2}P(\ell;\chi')d\chi'\nonumber\\
B_{ijk}(\ell_1,\ell_2,\ell_3)&=&\int_0^{\chi}\frac{W_{i}(\chi')W_{j}(\chi')W_{k}(\chi')}{\chi'^4}B(\ell_1,\ell_2,\ell_3;\chi')d\chi',
\end{eqnarray}
where $i,j,k$ denote the redshift bin and $W_i(\chi)$ are weighting functions which depend on the quantity being correlated. For example, when correlating weak lensing this is the weighted lensing kernel $W_i(\chi)=\int_0^{\chi}W_L(\chi',\chi)f_i(\chi')d\chi'$ where $f_i(\chi)$ is the comoving galaxy distribution in the \emph{i}-th redshift bin.

\subsection{The 2-point GI self-calibration}\label{tpsc}

For the power spectrum, \cite{22} proposed the GI self-calibration technique to calculate and remove the GI correlation, quantified by $f^I\equiv C^{IG}/C^{GG}$, from the angular cross-correlation power spectrum between galaxy ellipticity ($\kappa^s$) in the \emph{i}-th and \emph{j}-th redshift bins. We confirm and summarise the important components of this technique here, as it is a necessary component in our own development of the 3-point GGI self-calibration. We also present independent performance estimates calculated using the linear alignment model for intrinsic alignment of \cite{11} for the intrinsic alignment as described in Sec. \ref{error}.

The following three observable correlations in a lensing survey between galaxy surface density and convergence for $i<j$ are necessary to the GI self-calibration technique:

\begin{eqnarray}
C^{(1)}_{ij}(\ell) &\approx& C^{GG}_{ij}(\ell)+C^{IG}_{ij}(\ell),\nonumber\\
C^{(2)}_{ii}(\ell) &=& C^{gG}_{ii}(\ell)+C^{gI}_{ii}(\ell),\nonumber\\
C^{(3)}_{ii}(\ell) &=& C^{gg}_{ii}(\ell).\label{eq:2obs}
\end{eqnarray}
\begin{figure}
\includegraphics[angle=270,scale=0.7]{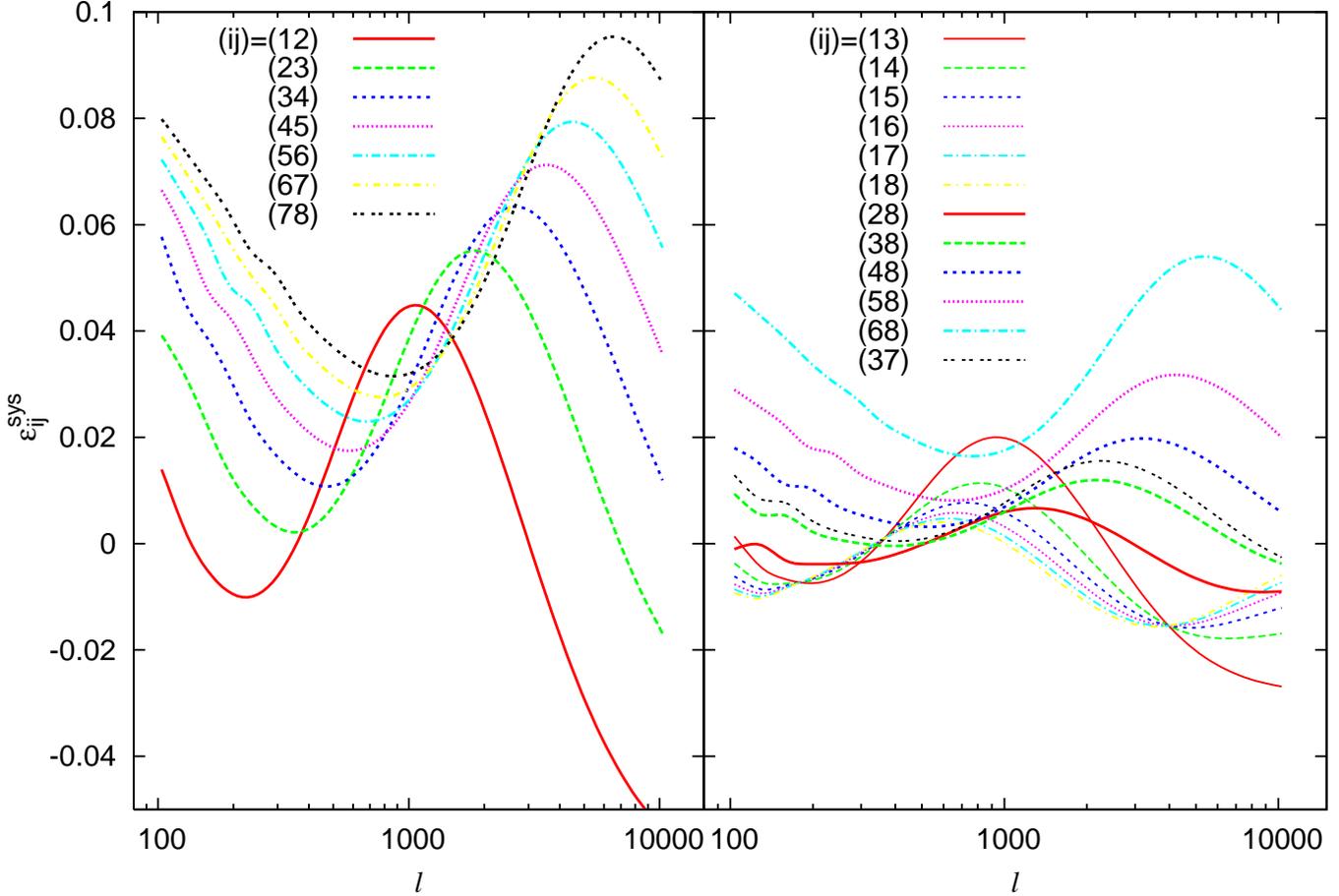}%
\caption{\label{fig:gi}The inaccuracy of the scaling relationship in Eq. \ref{eq:2scale} is quantified in Eq. \ref{eq:2eps} by $\epsilon^{sys}_{ij}$. This inaccuracy is the source of the dominant systematic error in the measurement of $C^{GG}_{ij}$ due to the GI self-calibration technique. Left: $\epsilon^{sys}_{ij}$ is plotted for adjacent redshift bins, where the stronger dependence of the lensing kernel on redshift causes a significantly higher inaccuracy. Right: $\epsilon^{sys}_{ij}$ is plotted for redshift bins of varying distance from each other. As expected, the inaccuracy for these bin choices is generally less than for three adjacent choices. Using the linear alignment model for intrinsic alignment of \protect\cite{11}, we find that our result is consistent with the toy model of \protect\cite{22} which gives the smallest systematic error. We thus expect a suppression of the GI intrinsic alignment contamination by at least a factor of $10$ for adjacent bins and up to a factor of $50$ for other bin pairs. These results are insensitive to the original GI contamination, such that for any $f^{thresh}_{ij}<f^I_{ij}<1$, the GI self-calibration will reduce the GI contamination down to survey limits or by a factor of $10$ or greater, whichever is less.}
\end{figure}
$C^{\alpha\beta}_{ij}$ is the angular cross-correlation power spectrum between quantity $\alpha$ in the \emph{i}-th redshift bin and $\beta$ in the \emph{j}-th redshift bin.  $\alpha,\beta\in G,I,g$, where \emph{G} indicates gravitational lensing ($\kappa$), \emph{I} the correlated galaxy intrinsic alignment ($\kappa^I$) and \emph{g} the galaxy number density distribution ($\delta^{\Sigma}$) in the corresponding redshift bin. Thus we denote the GI power spectrum $C^{IG}_{ij}$ in order to preserve the association of each quantity \emph{G} or \emph{I} to its redshift bin. By requiring $i<j$ with sufficient photo-z accuracy, $C^{GI}_{ij}$ and $C^{II}_{ij}$ provide negligible contribution to the lensing signal and are neglected in $C^{(1)}_{ij}$. From these observables, $C^{IG}_{ij}(\ell)$ is expressed by using a deterministic galaxy bias $b_1^i(\ell)$ (see Appendix A of \cite{22}) through the scaling relation

\begin{equation}
C^{IG}_{ij}(\ell)\approx \frac{W_{ij}}{b_1^i(\ell)\Pi_{ii}}C^{Ig}_{ii}(\ell),\label{eq:2scale}
\end{equation}
where $W_{ij}\equiv\int_0^{\infty}d\chi_L\int_0^{\infty}d\chi_G[W_L(\chi_L,\chi_G)f_i(\chi_L)f_j(\chi_G)]$ and $\Pi_{ii}=\int_0^{\infty} f_i^2(\chi)d\chi$. $f_i(\chi)=n_i(z)dz/d\chi$ is the true comoving distance distribution of galaxies in the \emph{i}-th redshift bin and $W_L(\chi_L,\chi_G)$ is the lensing kernel. Using the effects of lensing geometry, $C^{Ig}_{ii}(\ell)$ is isolated from the second observable $C^{(2)}_{ii}(\ell)$ (see Appendix B of \cite{22}) using the estimator

\begin{equation}
\hat{C}^{Ig}_{ii}(\ell)=\frac{C^{(2)}_{ii}|_S(\ell)-Q_2(\ell)C^{(2)}_{ii}(\ell)}{1-Q_2(\ell)},\label{eq:2cig}
\end{equation}
with $Q_2(\ell)\equiv C^{gG}_{ii}|_S(\ell)/C^{gG}_{ii}(\ell)$. The subscript \emph{S} denotes the correlation between only those pairs with $z_G^P<z_g^P$. $Q_2(\ell)$ then measures the relative suppression of the gG signal due to the orientation dependence of the lensing geometry, where $C^{gG}_{ii}|_S(\ell)\ll C^{gG}_{ii}(\ell)$. For spectroscopic (true) redshifts, $C^{gG}_{ii}|_S(\ell)=0$ and the estimator $\hat{C}^{Ig}_{ii}(\ell)$ is simply $C^{(2)}_{ii}|_S(\ell)=C^{gI}_{ii}(\ell)$, since the gI correlation is independent of orientation. The measurement error (see Appendix C of \cite{22}) in this estimator is 

\begin{eqnarray}
\Delta C^{Ig}_{ii}&=&\frac{1}{2\ell\Delta\ell f_{sky}}\bigg\{C^{gg}_{ii}C^{GG}_{ii}+\left(1+\frac{1}{3(1-Q_2)^2}\right)\left[C^{gg}_{ii}C^{gg,N}_{ii}+C^{gg,N}_{ii}(C^{GG}_{ii}+C^{II}_{ii})\right]\nonumber\\
&&+\left(1+\frac{1}{(1-Q_2)^2}\right)C^{gg,N}_{ii}C^{GG,N}_{ii}\bigg\}.
\end{eqnarray}

The fractional error on $C^{IG}_{ij}$ (and thus the residual statistical error in the measurement of $C^{GG}_{ij}$) due to $\Delta C^{Ig}_{ii}$ is

\begin{equation}
\Delta f_{ij}=\left(\frac{W_{ij}}{b_1^i(\ell)\Pi_{ii}}\right)\frac{\Delta C^{Ig}_{ii}}{C^{GG}_{ij}}.
\end{equation}
This is also the threshold contamination $f^I_{ij}=f^{thresh}_{ij}$ at which the GI self-calibration will function at S/N=1. Similarly, the scaling relation in Eq. \ref{eq:2scale} is not exact, and its accuracy is quantified by

\begin{equation}
\epsilon^{sys}_{ij}=\frac{b_1^i(\ell)\Pi_{ii}}{W_{ij}}\frac{C^{IG}_{ij}(\ell)}{C^{Ig}_{ij}(\ell)}-1.\label{eq:2eps}
\end{equation}
This introduces a residual systematic error in the measurement of $C^{GG}_{ij}$ of $\delta f_{ij}=\epsilon^{sys}_{ij}f^I_{ij}$ . 

The third observable $C^{(3)}_{ii}(\ell)$ can be used to calculate the galaxy bias. This gives a result for $C^{IG}_{ij}(\ell)$, which can then be removed from the first observable $C^{(1)}_{ij}(\ell)$. Our notation above, which we will use throughout this paper, is slightly different from the original notation of \cite{22} in order to be compatible with the GGI self-calibration. We denote analagous quantities in the 2- and 3-point self-calibration by the same variable, differentiated by the number of its indices.

The GI self-calibration technique converts a systematic intrinsic alignment contamination $f^I_{ij}$ into a residual statistical error $\Delta f_{ij}<f^I_{ij}$ which is insensitive to the intrinsic alignment contamination. We find good agreement with Zhang's estimation of the performance of the GI self-calibration, based on independent calculations of the 2-point errors following the methods described in Sec. \ref{error}. Figure \ref{fig:gi} shows the residual systematic ($\delta f_{ij}$) error in the measurement of $C^{GG}$ with the GI self-calibration technique. We use the linear alignment model for intrinsic alignment of \cite{11} and find that our result is consistent with the toy model of \cite{22} which gives the smallest systematic error. We thus expect for an LSST-like lensing survey a suppression of the GI intrinsic alignment contamination by at least a factor of 10 for adjacent bins and up to a factor of 50 for other bin pairs. These results are insensitive to the original GI contamination, such that for any $f^{thresh}_{ij}<f^I_{ij}<1$, the GI self-calibration will reduce the GI contamination down to survey limits or by a factor of 10 or greater, whichever is less.

\section{3-point GGI Self-Calibration}\label{sc}

There are several sets of correlations between the observed galaxy surface density and convergence which can be constructed for galaxy triplets. Only three of these observed correlations are needed for the GGI self-calibration technique. The first is the angular cross-correlation bispectrum between galaxy ellipticity ($\kappa^s$) in the \emph{i}-th, \emph{j}-th and \emph{k}-th redshift bin

\begin{equation}
B^{(1)}_{ijk}(\ell_1,\ell_2,\ell_3)=B^{GGG}_{ijk}(\ell_1,\ell_2,\ell_3)+B^{IGG}_{ijk}(\ell_1,\ell_2,\ell_3)+(\textrm{2 perm.})+B^{IIG}_{ijk}(\ell_1,\ell_2,\ell_3)+(\textrm{2 perm.})+B^{III}_{ijk}(\ell_1,\ell_2,\ell_3)\label{eq:onea}.
\end{equation}
$B^{\alpha\beta\gamma}_{ijk}$ is the angular cross-correlation bispectrum between quantity $\alpha$ in the \emph{i}-th redshift bin, $\beta$ in the \emph{j}-th redshift bin and $\gamma$ in the \emph{k}-th redshift bin. $\alpha,\beta,\gamma\in G,I,g$, where \emph{G} indicates gravitational lensing ($\kappa$), \emph{I} the correlated galaxy intrinsic alignment ($\kappa^I$) and \emph{g} the galaxy number density distribution ($\delta^{\Sigma}$) in the corresponding redshift bin. Unless catastrophic photo-z errors overwhelm the data, we can safely neglect the correlations GII and III which require spatially close galaxies by selecting galaxy triplets where $i<j<k$. Under this requirement, we also have $B^{IGG}_{ijk}\gg B^{GGI}_{ijk},B^{GIG}_{ijk}$ due to the lensing geometry. We then have for $i<j<k$,

\begin{equation}
B^{(1)}_{ijk}(\ell_1,\ell_2,\ell_3)\approx B^{GGG}_{ijk}(\ell_1,\ell_2,\ell_3)+B^{IGG}_{ijk}(\ell_1,\ell_2,\ell_3).\label{eq:one}
\end{equation}
Thus the dominant intrinsic alignment contamination is from the GGI bispectrum $B^{IGG}_{ijk}(\ell_1,\ell_2,\ell_3)$ ($i<j<k$), which the GGI self-calibration technique seeks to calculate and remove. Here we denote the GGI bispectrum $B^{IGG}_{ijk}$ in order to preserve the association of each quantity \emph{G} or \emph{I} to its redshift bin, as for the 2-point GI power spectrum.

The second correlation is measured in the angular cross-correlation bispectrum between convergence ($\kappa^s$) in the \emph{i}-th redshift bin and galaxy density ($\delta^{\Sigma}$) in the \emph{j}-th and \emph{k}-th redshift bins. Of interest to the self-calibration is the case where $i=j=k$, and we have

\begin{equation}
B^{(2)}_{iii}(\ell_1,\ell_2,\ell_3)=B^{Ggg}_{iii}(\ell_1,\ell_2,\ell_3)+B^{Igg}_{iii}(\ell_1,\ell_2,\ell_3).\label{eq:two}
\end{equation}
This correlation contributes further information about the intrinsic alignment of galaxies

The final correlation of interest is measured in the angular cross-correlation bispectrum between galaxy density ($\delta^{\Sigma}$) in the \emph{i}-th, \emph{j}-th and \emph{k}-th redshift bins when $i=j=k$, giving

\begin{equation}
B^{(3)}_{iii}(\ell_1,\ell_2,\ell_3)=B^{ggg}_{iii}(\ell_1,\ell_2,\ell_3).\label{eq:three}
\end{equation}

We also require for the GGI self-calibration those observables in Eq. \ref{eq:2obs} for the GI self-calibration. It is important to note that we have thus far neglected the contribution of magnification bias to these measurements. This will be further discussed and justified for the 3-point measurements in Sec. \ref{mag} and was discussed and shown to be negligible for the GI self-calibration by \cite{22}. There is also a non-Gaussian contribution to the observed bispectra. We briefly discuss the impact of this non-Gaussianity on the self-calibration technique in Sec. \ref{bias}, but otherwise leave discussion and calculation of this non-Gaussian contribution to the bispectrum to other works.

Our GGI self-calibration technique will calculate and remove the GGI contamination in Eq. \ref{eq:one} by using the measurements from Eqs. \ref{eq:two} \& \ref{eq:three}, which are both available in the same lensing survey. We express the fractional contamination to the lensing signal by the correlated intrinsic alignment as

\begin{equation}
f^I_{ijk}(\ell_1,\ell_2,\ell_3)\equiv \frac{B^{IGG}_{ijk}(\ell_1,\ell_2,\ell_3)}{B^{GGG}_{ijk}(\ell_1,\ell_2,\ell_3)}.
\end{equation}
For the self-calibration to work, the contamination $f^I_{ijk}(\ell_1,\ell_2,\ell_3)$ must be sufficiently large as to contribute a detectable $B^{Igg}_{iii}(\ell_1,\ell_2,\ell_3)$ at the corresponding $\ell$ bins in $B^{(2)}_{iii}(\ell_1,\ell_2,\ell_3)$. We denote this threshold $f^{thresh}_{ijk}$. When $f^I_{ijk}\ge f^{thresh}_{ijk}$, the GGI self-calibration can be applied to reduce the GGI contamination. The residual error after the GGI self-calibration will be expressed as a residual fractional error on the lensing measurement. In our notation, we differentiate $\Delta f_{ijk}$ as statistical error and $\delta f_{ijk}$ as systematic error. The performance of the GGI self-calibration will then be quantified by the parameters $f^{thresh}_{ijk}$, $\Delta f_{ijk}$ and $\delta f_{ijk}$, which are discussed and calculated in Sec. \ref{error}.

\subsection{Relationship between $B^{IGG}_{ijk}$ and $B^{Igg}_{iii}$}\label{scaling}

The first step in the GGI self-calibration is to determine the relationship between $B^{IGG}_{ijk}$ and $B^{Igg}_{iii}$. Under the Limber approximation, the 2D GGI angular cross-correlation bispectrum between the \emph{i}-th, \emph{j}-th and \emph{k}-th redshift bins is related to the 3D matter-matter-galaxy intrinsic alignment bispectrum by

\begin{equation}
B^{IGG}_{ijk}(\ell_1,\ell_2,\ell_3)=\int_0^{\infty}\frac{f_{i}(\chi)W_{j}(\chi)W_{k}(\chi)}{\chi^{4}}B_{\delta\delta\gamma^I}\left(k_1=\frac{\ell_1}{\chi},k_2=\frac{\ell_2}{\chi},k_3=\frac{\ell_3}{\chi};\chi\right)d\chi,\label{eq:scale1}
\end{equation}
where 

\begin{equation}
W_i(\chi_L)=\int_0^{\infty}W_L(\chi_L,\chi_G)f_i(\chi_G)d\chi_G.
\end{equation}
The integral runs from zero to $\infty$ in order to take into account the photo-z error. We again denote the GGI bispectrum $B^{IGG}_{ijk}$ in order to preserve the association of each quantity \emph{G} or \emph{I} to its redshift bin and will continue this convention throughout the paper. Similarly, the 2D ggI angular auto-correlation bispectrum is related to the 3D galaxy-galaxy-galaxy intrinsic alignment bispectrum by
\begin{equation}
B^{Igg}_{iii}(\ell_1,\ell_2,\ell_3)=\int_0^{\infty}\frac{f^3_{i}(\chi)}{\chi^{4}}B_{gg\gamma^I}\left(k_1=\frac{\ell_1}{\chi},k_2=\frac{\ell_2}{\chi},k_3=\frac{\ell_3}{\chi};\chi\right)d\chi.\label{eq:scale2}
\end{equation}
We will adopt a deterministic galaxy bias $b_{g,k}$ \citep{26} such that the smoothed galaxy density is a function of matter density expressed as
\begin{equation}
\delta_g(\bm{x};\chi)=b_{g,1}(\chi)\delta_m(\bm{x};\chi)+\frac{b_{g,2}(\chi)}{2}\delta_m^2(\bm{x};\chi)+O(\delta^3).\label{eq:density} 
\end{equation}
The first term $b_{g,1}$ is the linear galaxy bias (as used by \cite{22} for the 2-point correlations). The second term represents the first order non-linear contribution. $b_{g,2}$ is typically found to be negative and $\le b_{g,1}$ \citep{27}. Unlike in the 2-point case, it is insufficient to model the bias as simply scale dependent \citep{28}. Following the galaxy-galaxy-galaxy halo bispectrum derivation of \cite{28}, we use this expression of the galaxy density to relate $B^{IGG}_{\delta\delta\gamma^I}$ to $B^{Igg}_{gg\gamma^I}$. We neglect the portion of the bispectrum due to primordial non-Gaussianity and the trispectrum term, which contains further information about the non-Gaussianity. This is justified and discussed further in Sec. \ref{bias}. This results in the relationship
\begin{eqnarray}
B_{gg\gamma^I}(k_1,k_2,k_3;\chi)&=&b^2_{g,1}(\chi)B_{\delta\delta\gamma^I}(k_1,k_2,k_3;\chi)\nonumber\\
&&+b_{g,1}(\chi)b_{g,2}(\chi)\left[P_{\delta\gamma^I}(k_1;\chi)P_{\delta\delta}(k_2;\chi)+P_{\delta\delta}(k_2;\chi)P_{\delta\gamma^I}(k_3;\chi)+P_{\delta\gamma^I}(k_1;\chi)P_{\delta\gamma^I}(k_3;\chi)\right].\label{eq:bg}
\end{eqnarray}

If the galaxy bias changes slowly over the \emph{i}-th redshift bin with median comoving distance $\chi_i$, we can write to a good approximation $b^i_k=b_{g,k}(\chi_i)$. Substituting Eq. \ref{eq:bg} into Eq. \ref{eq:scale2}, we have
\begin{eqnarray}
B^{Igg}_{iii}(\ell_1,\ell_2,\ell_3)&=&\int_0^{\infty}\frac{f^3_{i}(\chi)}{\chi^{4}}\Big( (b^{i}_{1})^2B_{\delta\delta\gamma^I}(k_1,k_2,k_3;\chi)\nonumber\\
&&+b^{i}_{1}b^{i}_{2}\left[P_{\delta\gamma^I}(k_1;\chi)P_{\delta\delta}(k_2;\chi)+P_{\delta\delta}(k_2;\chi)P_{\delta\gamma^I}(k_3;\chi)+P_{\delta\gamma^I}(k_1;\chi)P_{\delta\gamma^I}(k_3;\chi)\right]\Big)d\chi.\label{eq:scale2a}
\end{eqnarray}
We can further approximate $B(k_1,k_2,k_3;\chi)\approx B(k_1,k_2,k_3;\chi_i)$ and $P(k;\chi)\approx P(k;\chi_i)$ in the limit where the comoving distance distribution of galaxies in the \emph{i}-th redshift bin is narrow. This leads to the following approximations of Eqs. \ref{eq:scale1} \& \ref{eq:scale2a},

\begin{equation}
B^{IGG}_{ijk}(\ell_1,\ell_2,\ell_3)\approx B_{\delta\delta\gamma^I}(k_1,k_2,k_3;\chi_i)\frac{W_{ijk}}{\chi_i^4},\label{eq:scale1b}
\end{equation}
and
\begin{eqnarray}
B^{Igg}_{iii}(\ell_1,\ell_2,\ell_3)&\approx & \frac{\Pi_{iii}}{\chi_i^4}\Big( (b^{i}_{1})^2B_{\delta\delta\gamma^I}(k_1,k_2,k_3;\chi_i)\nonumber\\
&&+b^{i}_{1}b^{i}_{2}\left[P_{\delta\gamma^I}(k_1;\chi_i)P_{\delta\delta}(k_2;\chi_i)+P_{\delta\delta}(k_2;\chi_i)P_{\delta\gamma^I}(k_3;\chi_i)+P_{\delta\gamma^I}(k_1;\chi_i)P_{\delta\gamma^I}(k_3;\chi_i)\right]\Big),\label{eq:scale2b}
\end{eqnarray}
where $W_{ijk}=\int_0^{\infty}f_i(\chi)W_j(\chi)W_k(\chi)d\chi$ and $\Pi_{iii}=\int_0^{\infty}f^3_i(\chi)d\chi$. From Eqs. \ref{eq:scale1b} \& \ref{eq:scale2b}, we have

\begin{eqnarray}
B^{IGG}_{ijk}(\ell_1,\ell_2,\ell_3)&\approx & \frac{W_{ijk}}{(b^{i}_{1})^2\Pi_{iii}}B^{Igg}_{iii}(\ell_1,\ell_2,\ell_3)-\frac{b^{i}_{2}W_{ijk}}{b^{i}_{1}\chi_i^4}\nonumber\\
&&\times\left[P_{\delta\gamma^I}(k_1;\chi_i)P_{\delta\delta}(k_2;\chi_i)+P_{\delta\delta}(k_2;\chi_i)P_{\delta\gamma^I}(k_3;\chi_i)+P_{\delta\gamma^I}(k_1;\chi_i)P_{\delta\gamma^I}(k_3;\chi_i)\right].\label{eq:scale3}
\end{eqnarray}

In order to express the 3D power spectra in Eq. \ref{eq:scale3} as 2D spectra, we will use the approximation made by Zhang $C^{Ig}_{ii}(\ell)\approx P_{\delta\gamma^I}(k;\chi_i)\frac{b^{i}_{1}\Pi_{ii}}{\chi^2_i}$ and the similar approximation $C^{GG}_{ii}(\ell)\approx P_{\delta\delta}(k;\chi_i)\frac{\omega_{ii}}{\chi^2_i}$, where $\omega_{ii}=\int_0^{\infty}W^2_i(\chi)d\chi$. Equation \ref{eq:scale3} is then

\begin{eqnarray}
B^{IGG}_{ijk}(\ell_1,\ell_2,\ell_3)&\approx & \frac{W_{ijk}}{(b^{i}_{1})^2\Pi_{iii}}B^{Igg}_{iii}(\ell_1,\ell_2,\ell_3)-\frac{b^{i}_{2}}{(b^{i}_{1})^2}\frac{W_{ijk}}{\omega_{ii}\Pi_{ii}}\nonumber\\
&&\times\left[C^{Ig}_{ii}(\ell_1)C^{GG}_{ii}(\ell_2)+C^{GG}_{ii}(\ell_2)C^{Ig}_{ii}(\ell_3)+\frac{\omega_{ii}}{b^i_1\Pi_{ii}}C^{Ig}_{ii}(\ell_1)C^{Ig}_{ii}(\ell_3)\right].\label{eq:scale4}
\end{eqnarray}

This relationship, while developed in the same way as for the GI self-calibration, is necessarily more complicated due to the inclusion of the non-linear galaxy bias and the presence of the GG correlation. Thus in order to apply this relationship, it is necessary to not only develop an estimator for $B^{Igg}_{iii}(\ell_1,\ell_2,\ell_3)$, which we describe in Sec. \ref{cigg}, but also to use the estimator $\hat{C}^{Ig}_{ii}(\ell)$ in Eq. \ref{eq:2cig} developed for the GI self-calibration and the resulting $C^{GG}_{ii}(\ell)$, as measured by the GI self-calibration \citep{22}. The GGI self-calibration technique is thus dependent upon the resulting measurements of the GI self-calibration technique.
\subsection{$B^{Igg}_{iii}$ Measurement}\label{cigg}
Information about the galaxy density-density-intrinsic ellipticity bispectrum, $B^{Igg}_{iii}$, is contained within the observable $B^{(2)}_{iii}=B^{Igg}_{iii}+B^{Ggg}_{iii}$. To measure it directly, we must first remove the contamination of $B^{Ggg}_{iii}$. For a spectroscopic galaxy sample, lensing geometry requires eliminating those triplets of galaxies where the redshift of the galaxy used to measure the ellipticity is lower than those used to measure galaxy number density. In this way, those triplets remaining have no contamination from $B^{Ggg}_{iii}$ and measure only $B^{Igg}_{iii}$.

In the case of a photo-z galaxy sample, this is not possible due to typically large photo-z error. Even for a photo-z bin with $\Delta z\rightarrow 0$, the photo-z error causes a true redshift distribution of width $\ge 2\sigma_P=0.1(1+z)$. In practice, photo-z bins are typically $\ge 0.2$. With such large errors, it is possible for galaxy triplets in the \emph{i}-th redshift bin to provide a measureable lensing contribution to $B^{Ggg}_{iii}$ even when requiring that the redshift of the galaxy used to measure the ellipticity is lower than those used to measure galaxy number density, except for the special cases where we limit to sufficiently low values the redshift or both the photo-z error and bin size. A more careful approach is thus required when separating $B^{Igg}_{iii}$ from $B^{Ggg}_{iii}$ for a general photo-z galaxy sample.

We apply the approach used by Zhang for the power spectrum $C^{(2)}_{ii}$ to the bispectrum $B^{(2)}_{iii}$, wherein we consider the orientation dependence of the two components. We will first define a redshift for each galaxy in the triplet: $z_{G/I}$ for the galaxy used in the lensing/intrinsic alignment measurement and $z_g$, $z_{g'}$ for the two galaxies used in the number density measurement. The ggI correlation is independent of the relative position of the three galaxies. For example, the correlations with $z_I< z_g< z_{g'}$, $z_g< z_I< z_{g'}$ or $z_g< z_{g'}< z_I$ are statistically identical when the sides of the triangle are fixed. However, the Ggg correlation does depend on the relative position of the three galaxies. Due to the lensing geometry dependence, the correlation with $z_G < z_g, z_{g'}$ is statistically smaller than other orientations.

\begin{figure}
\includegraphics[angle=270,scale=0.50]{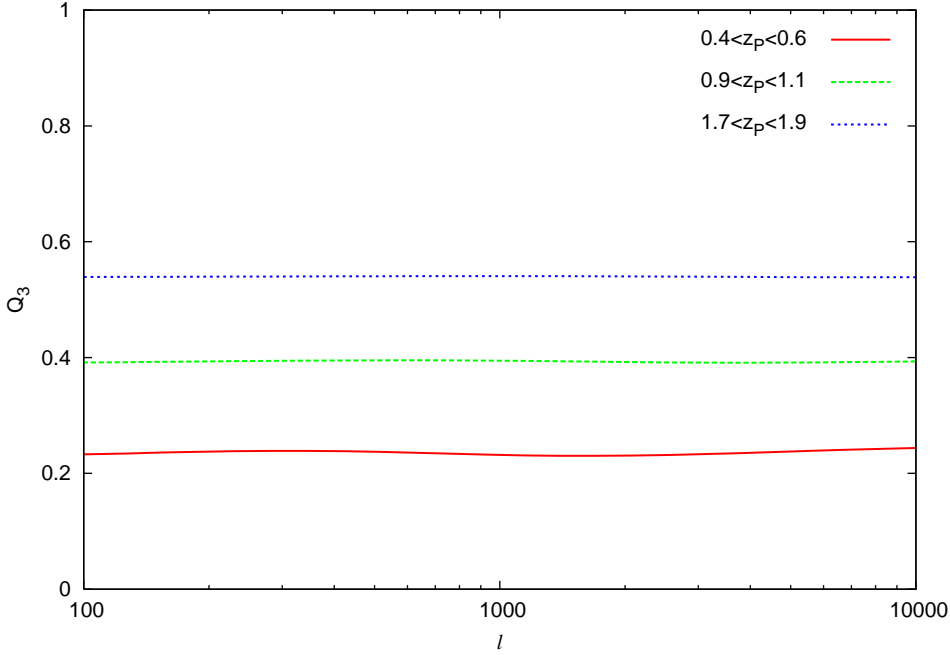}%
\caption{\label{fig:q}The behavior of $Q_3(\ell_1,\ell_2,\ell_3)\equiv B^{Ggg}_{iii}|_S(\ell_1,\ell_2,\ell_3)/B^{Ggg}_{iii}(\ell_1,\ell_2,\ell_3)$ for equilateral triangles ($\ell=\ell_1=\ell_2=\ell_3$) over three redshift bins spanning the survey range. Similar to the $2$-point case, the suppression is dependent on the redshift bin chosen, increasing with redshift due to increased photo-z error at higher redshift, but is largely scale independent due to being the ratio of two bispectra. For this reason there is also little dependence on triangle shape. Generally, $Q\approx 0.4$, and the significant deviation from unity ensures that the estimator $\hat{B}^{Igg}_{iii}$ is valid for lensing surveys of interest.}
\end{figure}

This dependence provides two observables from $B^{(2)}_{iii}$. The first is $B^{(2)}_{iii}$, where all triplets are weighted equally. The second is $B^{(2)}_{iii}|_S$, which counts only those triplets with $z_G< z_g, z_{g'}$. This weighting is denoted by the subscript 'S'. From our previous discussion, we then have $B^{Igg}_{iii}=B^{Igg}_{iii}|_S$ and $B^{Ggg}_{iii}> B^{Ggg}_{iii}|_S$. We now define the ratio

\begin{equation}
Q_3(\ell_1,\ell_2,\ell_3)\equiv\frac{B^{Ggg}_{iii}|_S(\ell_1,\ell_2,\ell_3)}{B^{Ggg}_{iii}(\ell_1,\ell_2,\ell_3)},\label{eq:qfact}
\end{equation}
where we have explicitly included the $\ell$-dependence which had been neglected previously in this section. This ratio describes the suppression of the signal due to the weighting of triplets described previously. By definition $0< Q_3< 1$, with $Q_3=0$ if the photo-z is perfectly accurate and $Q_3=1$ if the photo-z has no correlation to the true redshift. $Q_3$ is calculated using the galaxy redshift distribution, which is discussed in Sec. \ref{qfact}.

We now define an estimator for $B^{Igg}_{iii}$ (that we denote $\hat{B}^{Igg}_{iii}$) in terms of $Q_3(\ell_1,\ell_2,\ell_3)$ and the two observables

\begin{eqnarray}
B^{(2)}_{iii}(\ell_1,\ell_2,\ell_3)&=&B^{Igg}_{iii}(\ell_1,\ell_2,\ell_3)+B^{Ggg}_{iii}(\ell_1,\ell_2,\ell_3),\nonumber\\
B^{(2)}_{iii}|_S(\ell_1,\ell_2,\ell_3)&=&B^{Igg}_{iii}(\ell_1,\ell_2,\ell_3)+B^{Ggg}_{iii}|_S(\ell_1,\ell_2,\ell_3)\label{eq:b2s}.
\end{eqnarray}
This estimator is

\begin{equation}
\hat{B}^{Igg}_{iii}(\ell_1,\ell_2,\ell_3)=\frac{B^{(2)}_{iii}|_S(\ell_1,\ell_2,\ell_3)-Q_3(\ell_1,\ell_2,\ell_3)B^{(2)}_{iii}(\ell_1,\ell_2,\ell_3)}{1-Q_3(\ell_1,\ell_2,\ell_3)}\label{eq:cig}.
\end{equation}

As expected, when $Q_3=0$ this gives $\hat{B}^{Igg}_{iii}=B^{(2)}_{iii}|_S$ as for a spectroscopic galaxy sample with no photo-z error. However, $Q_3$ must not approach unity, where $\hat{B}^{Igg}_{iii}$ is singular. For the LSST-like survey described in Sec. \ref{survey}, we calculate $Q_3$ for various redshift bins following the procedure described in Sec. \ref{qfact}. This result is given in Fig. 2 for equilateral triangles, where we find $Q_3\approx 0.4$ and in general that $Q_3$ should deviate significantly from unity. The estimator $\hat{B}^{Igg}_{iii}$ is thus expected to be applicable in any typical lensing survey.

\subsection{Evaluating $Q_3(\ell_1,\ell_2,\ell_3)$}\label{qfact}

In order to evaluate the ratio $Q_3$ in Eq. \ref{eq:qfact}, we will begin from the real space angular correlation function $w^{Ggg'}\left(\theta_1,\theta_2,\theta_3;z^P_G,z^P_g,z^P_{g'}\right)$ between shear at photo-z $z^P_G$ and galaxy density at photo-z $z^P_g$ and $z^P_{g'}$. The average correlation over the distribution of galaxies in the \emph{i}-th redshift bin is

\begin{eqnarray}
w^{Ggg'}_{iii}(\theta_1,\theta_2,\theta_3)&=&\int_idz^P_G\int_idz^P_g\int_idz^P_{g'}w^{Ggg'}\left(\theta_1,\theta_2,\theta_3;z^P_G,z^P_g,z^P_{g'}\right)n_i^P(z^P_G)n_i^P(z^P_g)n_i^P(z^P_{g'})\nonumber\\
&=&\int_idz^P_G\int_idz^P_g\int_idz^P_{g'}\int_0^{\infty}dz_G\int_0^{\infty}dz_g\int_0^{\infty}dz_{g'}w^{Ggg'}(\theta_1,\theta_2,\theta_3;z_G,z_g,z_{g'})\nonumber\\
&&\times p(z_G|z_G^P)p(z_g|z_g^P)p(z_{g'}|z_{g'}^P)n_i^P(z^P_G)n_i^P(z^P_g)n_i^P(z^P_{g'}),\label{eq:q1}
\end{eqnarray}
where we have used the shorthand $\int_i=\int_{\bar{z}_i-\Delta z_i/2}^{\bar{z}_i+\Delta z_i/2}$ to represent integration over the \emph{i}-th redshift bin. In terms of the ensemble average $\langle \cdots \rangle$, which is in practice an average over $\theta'$, the angular real space correlation function is

\begin{equation}
w^{Ggg'}(\theta_1,\theta_2,\theta_3;z_G,z_g,z_{g'})=\int_0^{\infty}dz_L\langle\delta_m(\theta';z_G)\delta_g(\theta'+\theta_2;z_g)\delta_g(\theta'+\theta_3;z_{g'})\rangle W_L(z_L,z_G)\delta^D(\theta_1+\theta_2+\theta_3),
\end{equation}
where $\delta^D(\theta_1+\theta_2+\theta_3)$ ensures that the three vectors form a triangle. We can now write Eq. \ref{eq:q1} as

\begin{eqnarray}
w^{Ggg'}_{iii}(\theta_1,\theta_2,\theta_3)&=&\int_idz^P_G\int_idz^P_g\int_idz^P_{g'}\int_0^{\infty}dz_G\int_0^{\infty}dz_g\int_0^{\infty}dz_{g'}\int_0^{\infty}dz_L\langle\delta_m(\theta';z_G)\delta_g(\theta'+\theta_2;z_g)\delta_g(\theta'+\theta_3;z_{g'})\rangle\nonumber\\
&&\times\delta^D(\theta_1+\theta_2+\theta_3)W_L(z_L,z_G)p(z_G|z_G^P)p(z_g|z_g^P)p(z_{g'}|z_{g'}^P)n_i^P(z^P_G)n_i^P(z^P_g)n_i^P(z^P_{g'})\nonumber\\
&&=\int_0^{\infty}dz_L\int_0^{\infty}dz_g\int_0^{\infty}dz_{g'}\langle\delta_m(\theta';z_G)\delta_g(\theta'+\theta_2;z_g)\delta_g(\theta'+\theta_3;z_{g'})\rangle\delta^D(\theta_1+\theta_2+\theta_3)\nonumber\\
&&\times W_i(z_L)n_i(z_g)n_i(z_{g'}).\label{eq:q2}
\end{eqnarray}

The second correlation function needed is identical to Eq. \ref{eq:q2}, but takes the average over all triplets such that $z^P_G< z^P_g, z^P_{g'}$,

\begin{eqnarray}
w^{Ggg'}_{iii}|_S(\theta_1,\theta_2,\theta_3)&=&\int_0^{\infty}dz_L\int_0^{\infty}dz_g\int_0^{\infty}dz_{g'}\langle\delta_m(\theta';z_G)\delta_g(\theta'+\theta_2;z_g)\delta_g(\theta'+\theta_3;z_{g'})\rangle\delta^D(\theta_1+\theta_2+\theta_3)\nonumber\\
&& \times W_i(z_L)n_i(z_g)n_i(z_{g'})\eta(z_L,z_g,z_{g'}).\label{eq:q3}
\end{eqnarray}
We have used here 

\begin{equation}
\eta(z_L,z_g,z_{g'})=\frac{3\int_i dz^P_G\int_i dz^P_g\int_i dz^P_{g'}\int_0^{\infty}dz_G W_L(z_L,z_G)p(z_G|z_G^P)p(z_g|z_g^P)p(z_{g'}|z_{g'}^P)n_i^P(z^P_G)n_i^P(z^P_g)n_i^P(z^P_{g'})S(z^P_G,z^P_g,z^P_{g'})}{\int_i dz^P_G\int_i dz^P_g\int_i dz^P_{g'}\int_0^{\infty}dz_G W_L(z_L,z_G)p(z_G|z_G^P)p(z_g|z_g^P)p(z_{g'}|z_{g'}^P)n_i^P(z^P_G)n_i^P(z^P_g)n_i^P(z^P_{g'})},\label{eq:eta}
\end{equation}
where $S(z^P_G,z^P_g,z^P_{g'})=1$ if $z^P_G< z^P_g, z^P_{g'}$ and is zero otherwise. Since $S(z^P_G,z^P_g,z^P_{g'})$ allows only $1/3$ of the integral to survive, $\eta(z_L,z_g,z_{g'})$ is normalised by a factor 3 in order to remove the suppression due to the selection function and measure only that due to the lensing geometry. This is demonstrated by the relation

\begin{equation}
\frac{\int_i dz^P_G\int_i dz^P_g\int_i dz^P_{g'}\int_0^{\infty}dz_G p(Z_G|z_G^P)p(Z_g|z_g^P)p(Z_{g'}|z_{g'}^P)n_i^P(z^P_G)n_i^P(z^P_g)n_i^P(z^P_{g'})S(z^P_G,z^P_g,z^P_{g'})}{\int_i dz^P_G\int_i dz^P_g\int_i dz^P_{g'}\int_0^{\infty}dz_G p(Z_G|z_G^P)p(Z_g|z_g^P)p(Z_{g'}|z_{g'}^P)n_i^P(z^P_G)n_i^P(z^P_g)n_i^P(z^P_{g'})}=\frac{1}{3}.
\end{equation}

We now take the Fourier transform of Eqs. \ref{eq:q2} \& \ref{eq:q3} to find the bispectra $B^{Ggg}_{iii}$ and $B^{Ggg}_{iii}|_S$, respectively. Again following the Limber approximation, with dominant correlation at $z_L=z_g=z_{g'}$, we have

\begin{equation}
B^{Ggg}_{iii}(\ell_1,\ell_2,\ell_3)=\int_0^{\infty}B^{Ggg}(k_1,k_2,k_3;\chi)\frac{W_i(\chi)f^2_i(\chi)}{\chi^4}d\chi\label{eq:q4}
\end{equation}
and

\begin{equation}
B^{Ggg}_{iii}|_S(\ell_1,\ell_2,\ell_3)=\int_0^{\infty}B^{Ggg}(k_1,k_2,k_3;\chi)\frac{W_i(\chi)f^2_i(\chi)}{\chi^4(z)}\eta(\chi,\chi(z_g)=\chi,\chi(z_{g'})=\chi)d\chi.\label{eq:q5}
\end{equation}

The ratio $Q_3$ is now expressed directly through Eqs. \ref{eq:q4} \& \ref{eq:q5}. We can approximate $Q_3\approx\bar{\eta}_i$, where $\bar{\eta}_i$ is the mean value of $\eta$ across the \emph{i}-th redshift bin, since the integrals differ only by a factor $\eta$. $\eta$ has the same dependence as $Q_3$ on the relative contribution to the Ggg correlation from triplets with $z^P_G< z^P_g, z^P_{g'}$ compared to triplets with other relative orientations. In the limit where photo-z error dominates, $\sigma_P\gg\Delta z$, and there is no suppression of the contribution to the Ggg correlation by the selection function, so $\eta,Q_3\rightarrow 1$. In this limit, the estimator $\hat{B}^{Igg}_{iii}$ becomes singular and $B^{Igg}_{iii}$ can no longer be differentiated from $B^{Ggg}_{iii}$. In the opposite limit, where $\sigma_p\ll\Delta z$, the selection function suppresses all contribution to the Ggg correlation and $\eta,Q_3\rightarrow 0$, where our estimator mirrors the extraction method for $B^{Igg}_{iii}$ in spectroscopic galaxy samples. 
\section{Performance of the GGI Self-Calibration}\label{error}
In order to evaluate the statistical and systematic errors in the GGI self-calibration, we calculate directly the power spectra and bispectra through the Limber approximation according to the anticipated survey parameters discussed in Sec. \ref{survey}. For the bispectra, we employ the fitting formula of \cite{30} for the 3D matter density bispectrum due to non-linear clustering. We modify this as described in Sec. \ref{scaling} for the 3D galaxy bispectrum, using values for the galaxy bias of $b^i_1=1.0$ and $b^i_2=-0.1$ \citep{31}. We include the intrinsic alignment correlations of $B^{IGG}$ and $B^{Igg}$ in a straightforward manner following the linear alignment model of \cite{11}, where $P_{\delta,\gamma^I}=-\frac{C_1\bar{\rho}}{D(z)(1+z)}P_{\delta}$. Like \cite{9a}, we extend this to the non-linear matter power spectrum for use in the fitting formula, where $C_1$ is estimated by comparison to Fig. 2 of \cite{11}.
\subsection{The estimation of $B^{Igg}_{iii}$}\label{ciggerror}
In order to quantify the accuracy of the estimator $\hat{B}^{Igg}_{iii}(\ell_1,\ell_2,\ell_3)$, we consider the contribution of measurement errors such as shot and shape noise in $\hat{B}^{(2)}_{iii}(\ell_1,\ell_2,\ell_3)$ which propagate into our measurement of $B^{Igg}_{iii}(\ell_1,\ell_2,\ell_3)$ through the estimator. We calculate the rms error for a given redshift bin, working in a pixel space with $N_P$ sufficiently fine and uniform pixels of photo-z with bin width $\Delta z$ and angular position with bin width $\Delta \ell$. Each pixel is associated with a photo-z $z^P_{\alpha}$, angular position $\theta_{\alpha}$, measured overdensity $\delta_{\alpha}+\delta^N_{\alpha}$ and measured `shear' $\kappa_{\alpha}+\kappa^I_{\alpha}+\kappa^N_{\alpha}$, where `N' represents the measurement noise. From Eq. \ref{eq:b2s}, we construct the pixel space angular bispectra

\begin{eqnarray}
B^{(2)}(\ell_1,\ell_2,\ell_3)&=&N_P^{-3}\sum_{\alpha\beta\gamma}[\delta_{\alpha}+\delta^N_{\alpha}][\delta_{\beta}+\delta^N_{\beta}][\kappa_{\gamma}+\kappa^I_{\gamma}+\kappa^N_{\gamma}]\exp[i(\bm{\ell_1}\cdot\bm{\theta_{\alpha}}+\bm{\ell_2}\cdot\bm{\theta_{\beta}}+\bm{\ell_3}\cdot\bm{\theta_{\gamma}})],\nonumber\\
B^{(2)}|_S(\ell_1,\ell_2,\ell_3)&=&N_P^{-3}\sum_{\alpha\beta\gamma}[\delta_{\alpha}+\delta^N_{\alpha}][\delta_{\beta}+\delta^N_{\beta}][\kappa_{\gamma}+\kappa^I_{\gamma}+\kappa^N_{\gamma}]\exp[i(\bm{\ell_1}\cdot\bm{\theta_{\alpha}}+\bm{\ell_2}\cdot\bm{\theta_{\beta}}+\bm{\ell_3}\cdot\bm{\theta_{\gamma}})]S_{\alpha\beta\gamma}.\label{eq:ester1}
\end{eqnarray}
$S_{\alpha\beta\gamma}=1$ when $z^P_{\alpha},z^P_{\beta}>z^P_{\gamma}$ and is zero otherwise. Thus in the limit $N_P\gg 1$, $\sum_{\alpha\beta\gamma}S_{\alpha\beta\gamma}=N_P^3/3$ and the average $\bar{S}_{\alpha\beta\gamma}=1/3$.

From our definition of the estimator in Eq. \ref{eq:cig}, we can construct the difference

\begin{eqnarray}
\hat{B}^{Igg}_{iii}-B^{Igg}_{iii}&=&\frac{1}{(1-Q_3)}N_P^{-3}\sum_{\alpha\beta\gamma}\exp[i(\bm{\ell_1}\cdot\bm{\theta_{\alpha}}+\bm{\ell_2}\cdot\bm{\theta_{\beta}}+\bm{\ell_3}\cdot\bm{\theta_{\gamma}})][(\delta_{\alpha}+\delta^N_{\alpha})(\delta_{\beta}+\delta^N_{\beta})(\kappa_{\gamma}+\kappa^I_{\gamma}+\kappa^N_{\gamma})(3S_{\alpha\beta\gamma}-Q_3)\nonumber\\
&&-(1-Q_3)\delta_{\alpha}\delta_{\beta}\kappa^I_{\gamma}]\nonumber\\
&=&\frac{1}{(1-Q_3)}N_P^{-3}\sum_{\alpha\beta\gamma}\exp[i(\bm{\ell_1}\cdot\bm{\theta_{\alpha}}+\bm{\ell_2}\cdot\bm{\theta_{\beta}}+\bm{\ell_3}\cdot\bm{\theta_{\gamma}})][(\delta_{\alpha}\delta^N_{\beta}+\delta^N_{\alpha}\delta_{\beta}+\delta^N_{\alpha}\delta^N_{\beta})(\kappa_{\gamma}+\kappa^I_{\gamma}+\kappa^N_{\gamma})+\delta_{\alpha}\delta_{\beta}(\kappa_{\gamma}+\kappa^N_{\gamma})]\nonumber\\
&&\times(3S_{\alpha\beta\gamma}-Q_3).
\end{eqnarray}
Here we have used $\bar{S}_{\alpha\beta\gamma}=1/3$ and that the ggI correlation doesn't depend on the relative position of the galaxy triplets. The rms error is 

\begin{eqnarray}
\left(\Delta B^{Igg}_{iii}\right)^2&=&\frac{1}{(1-Q_3)^2}N_P^{-6}\sum_{\alpha\beta\gamma}\sum_{\lambda\mu\nu}\exp[i(\bm{\ell_1}\cdot\bm{\theta_{\alpha}}+\bm{\ell_2}\cdot\bm{\theta_{\beta}}+\bm{\ell_3}\cdot\bm{\theta_{\gamma}})]\exp[i(\bm{\ell_1}\cdot\bm{\theta_{\lambda}}+\bm{\ell_2}\cdot\bm{\theta_{\mu}}+\bm{\ell_3}\cdot\bm{\theta_{\nu}})](3S_{\alpha\beta\gamma}-Q_3)\nonumber\\
&&\times(3S_{\lambda\mu\nu}-Q_3)\langle[(\delta_{\alpha}\delta^N_{\beta}+\delta^N_{\alpha}\delta_{\beta}+\delta^N_{\alpha}\delta^N_{\beta})(\kappa_{\gamma}+\kappa^I_{\gamma}+\kappa^N_{\gamma})+\delta_{\alpha}\delta_{\beta}(\kappa_{\gamma}+\kappa^N_{\gamma})]\nonumber\\
&&\times[(\delta_{\lambda}\delta^N_{\mu}+\delta^N_{\lambda}\delta_{\nu}+\delta^N_{\nu}\delta^N_{\mu})(\kappa_{\nu}+\kappa^I_{\nu}+\kappa^N_{\nu})+\delta_{\lambda}\delta_{\mu}(\kappa_{\nu}+\kappa^N_{\nu})]\rangle,
\end{eqnarray}
where $\langle\cdots\rangle$ is the ensemble average. The ensemble average is over 121 terms of the form $\langle ABCDEF\rangle$, $A,B,C,D,E,F \in \delta,\delta^N,\kappa,\kappa^I,\kappa^N$. To simplify this we apply Wick's theorem for the 6-point correlation,

\begin{equation}
\langle ABCDEF\rangle=\langle AB\rangle\langle CD\rangle\langle EF\rangle+\langle AB\rangle\langle CE\rangle\langle DF\rangle+(\textrm{14 perm.}).
\end{equation}

This results in 1815 products of three 2-point correlations, most of which are zero. Any correlation between signal and noise or dissimilar noise terms vanish. Due to the angular dependence of the correlations ($\langle A_{a}B_{b}\rangle=w_{AB}(\theta_{a}-\theta_{b})$), only those correlations with $\langle A_{a}B_{b}\rangle$ where \emph{a} $\in \alpha,\beta,\gamma$ and \emph{b} $\in \lambda,\mu,\nu$ are non-vanishing. This leaves 42 surviving products:

\begin{eqnarray}
\left(\Delta B^{Igg}_{iii}\right)^2&=&\frac{1}{\left(1-Q_3\right)^2}N_P^{-6}\sum_{\alpha\beta\gamma}\sum_{\lambda\mu\nu}\exp[i(\bm{\ell_1}\cdot\bm{\theta_{\alpha}}+\bm{\ell_2}\cdot\bm{\theta_{\beta}}+\bm{\ell_3}\cdot\bm{\theta_{\gamma}})]\exp[i(\bm{\ell_1}\cdot\bm{\theta_{\lambda}}+\bm{\ell_2}\cdot\bm{\theta_{\mu}}+\bm{\ell_3}\cdot\bm{\theta_{\nu}})]\left(3S_{\alpha\beta\gamma}-Q_3\right)\nonumber\\
&&\times\left(3S_{\lambda\mu\nu}-Q_3\right)\big[\langle\delta_{\alpha}\delta_{\lambda}\rangle\left[\left(\langle\delta_{\beta}\delta_{\mu}\rangle+\langle\delta^N_{\beta}\delta^N_{\mu}\rangle\right)\left(\langle\kappa_{\gamma}\kappa_{\nu}\rangle+\langle\kappa^N_{\gamma}\kappa^N_{\nu}\rangle\right)+\langle\delta^N_{\beta}\delta^N_{\mu}\rangle\langle \kappa^I_{\gamma}\kappa^I_{\nu}\rangle+\langle\kappa_{\gamma}\delta_{\mu}\rangle\langle\delta_{\beta}\kappa_{\nu}\rangle\right]\nonumber\\
&&+\langle\delta_{\alpha}\delta_{\mu}\rangle\left[\left(\langle\delta_{\beta}\delta_{\lambda}\rangle+\langle\delta^N_{\beta}\delta^N_{\lambda}\rangle\right)\left(\langle\kappa_{\gamma}\kappa_{\nu}\rangle+\langle\kappa^N_{\gamma}\kappa^N_{\nu}\rangle\right)+\langle\delta^N_{\beta}\delta^N_{\lambda}\rangle\langle \kappa^I_{\gamma}\kappa^I_{\nu}\rangle+\langle\kappa_{\gamma}\delta_{\lambda}\rangle\langle\delta_{\beta}\kappa_{\nu}\rangle\right]\nonumber\\
&&+\langle\delta^N_{\alpha}\delta^N_{\lambda}\rangle\left[\left(\langle\delta_{\beta}\delta_{\mu}\rangle+\langle\delta^N_{\beta}\delta^N_{\mu}\rangle\right)\left(\langle\kappa_{\gamma}\kappa_{\nu}\rangle+\langle\kappa^N_{\gamma}\kappa^N_{\nu}\rangle+\langle \kappa^I_{\gamma}\kappa^I_{\nu}\rangle\right)+\langle\left(\kappa_{\gamma}+I_{\gamma}\right)\delta_{\mu}\rangle\langle\delta_{\beta}\left(\kappa_{\nu}+\kappa^I_{\nu}\right)\rangle\right]\nonumber\\
&&+\langle\delta^N_{\alpha}\delta^N_{\mu}\rangle\left[\left(\langle\delta_{\beta}\delta_{\lambda}\rangle+\langle\delta^N_{\beta}\delta^N_{\lambda}\rangle\right)\left(\langle\kappa_{\gamma}\kappa_{\nu}\rangle+\langle\kappa^N_{\gamma}\kappa^N_{\nu}\rangle+\langle \kappa^I_{\gamma}\kappa^I_{\nu}\rangle\right)+\langle\left(\kappa_{\gamma}+\kappa^I_{\gamma}\right)\delta_{\lambda}\rangle\langle\delta_{\beta}\left(\kappa_{\nu}+I_{\nu}\right)\rangle\right]\nonumber\\
&&+\langle\delta_{\alpha}\kappa_{\nu}\rangle\left[\langle\delta_{\beta}\delta_{\mu}\rangle\langle\kappa_{\gamma}\delta_{\lambda}\rangle+\langle\delta_{\beta}\delta_{\lambda}\rangle\langle\kappa_{\gamma}\delta_{\mu}\rangle\right]+\langle\delta_{\alpha}\left(\kappa_{\nu}+\kappa^I_{\nu}\right)\rangle\left[\langle\delta^N_{\beta}\delta^N_{\mu}\rangle\langle\left(\kappa_{\gamma}+\kappa^I_{\gamma}\right)\delta_{\lambda}\rangle+\langle\delta^N_{\beta}\delta^N_{\lambda}\rangle\langle\left(\kappa_{\gamma}+\kappa^I_{\gamma}\right)\delta_{\mu}\rangle\right]\big].\label{eq:ester2}
\end{eqnarray}

Noises only correlate at zero lag ($\langle\delta^N_{\alpha}\delta^N_{\lambda}\rangle\propto \delta_{\alpha\lambda}$, $\langle\kappa^N_{\gamma}\kappa^N_{\nu}\rangle\propto \delta_{\gamma\nu}$), and the correlations $\langle\delta\delta\rangle$, $\langle\delta \kappa^I\rangle$, $\langle\kappa\kappa\rangle$ and $\langle \kappa^I\kappa^I\rangle$ depend only on separation, not on relative orientation of the galaxy pairs along the line-of-sight. However, $\langle\kappa\delta\rangle$ is dependent on the relative orientation along the line-of-sight and must be treated with care when evaluating Eq. \ref{eq:ester2}. In order to quantify this orientation dependence, we apply $Q_2$ such that

\begin{equation}
\langle\delta_{\alpha}\kappa_{\nu}\rangle\rightarrow\frac{1}{2} \left(\frac{S_{\alpha\nu}}{(1-Q_2)}+\frac{S_{\nu\alpha}}{Q_2}\right)\langle\delta_{\alpha}\kappa_{\nu}\rangle.\label{eq:sub}
\end{equation}

\begin{figure}
\includegraphics[angle=270,scale=0.7]{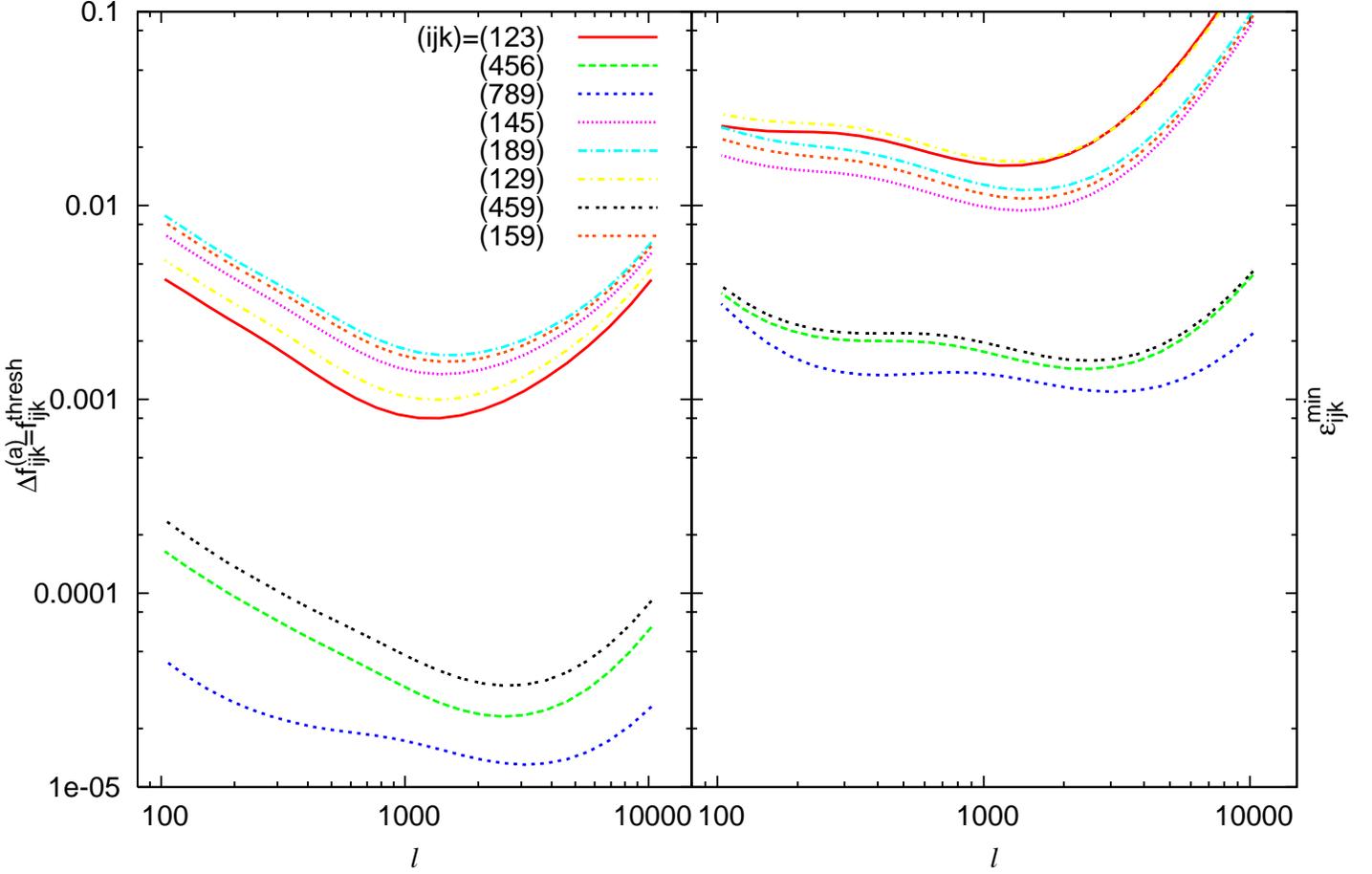}%
\caption{\label{fig:merror}Left: The residual statistical uncertainty $\Delta f^{(a)}_{ijk}$ in the $B^{IGG}_{ijk}$ measurement and threshold of intrinsic alignment contamination $f^{thresh}_{ijk}$ at which the GGI self-calibration technique can calculate and remove the intrinsic alignment contamination at S/N=1 are plotted for a variety of redshift bin combinations. Right: The minimum measurement error $\epsilon^{min}_{ijk}$ for $B^{GGG}_{ijk}$ is plotted for comparison. Both errors are plotted for equilateral triangles ($\ell=\ell_1=\ell_2=\ell_3$) and have a similar $\ell$ dependence, with the effects of shot noise taking over at large $\ell$. Generally, $\Delta f^{(a)}_{ijk}<\epsilon^{min}_{ijk}$, and is thus negligible. We expect this result to hold for non-equilateral triangles as well, but the use of the GGI self-calibration is limited by our understanding of non-Gaussian effects for very elongated triangle shapes, as discussed in Sec. \ref{bias}, and we leave discussion of its applicability for these very elongated triangle shapes to a future work.}
\end{figure}
We can now evaluate Eq. \ref{eq:ester2} analytically, taking the Fourier transform to find

\begin{eqnarray}
\left(\Delta B^{Igg}_{iii}\right)^2&=&2\Big\{\left(C^{GG}_{ii}C^{gg}_{ii}+bC^{GG,N}_{ii}C^{gg}_{ii}+2fC^{gG}_{ii}C^{gG}_{ii}\right)C^{gg}_{ii}\nonumber\\
&&+2\Big(a\left[\left(C^{GG}_{ii}+C^{II}_{ii}\right)C^{gg}_{ii}+C^{gI}_{ii}C^{gI}_{ii}\right]+dC^{GG,N}_{ii}C^{gg}_{ii}+gC^{gG}_{ii}C^{gG}_{ii}+2hC^{gG}_{ii}C^{gI}_{ii}\nonumber\\
&&+\left[c\left(C^{GG}_{ii}+C^{II}_{ii}\right)+eC^{GG,N}_{ii}\right]C^{gg,N}_{ii}\Big)C^{gg,N}_{ii}\Big\}.\label{eq:ester3}
\end{eqnarray}
The details of this calculation and the coefficients $a-h$ are found in the Appendix.

The final rms error $\Delta B^{Igg}_{iii}$ evaluated for a given triangle with bin width $\Delta \ell$ is then given by

\begin{eqnarray}
\left(\Delta B^{Igg}_{iii}\right)^2&=&\frac{4\pi^2}{\ell_1\ell_2\ell_3\Delta\ell_1\Delta\ell_2\Delta\ell_3f_{sky}}\Big\{\left(C^{GG}_{ii}C^{gg}_{ii}+bC^{GG,N}_{ii}C^{gg}_{ii}+2fC^{gG}_{ii}C^{gG}_{ii}\right)C^{gg}_{ii}\nonumber\\
&&+2\Big(a\left[\left(C^{GG}_{ii}+C^{II}_{ii}\right)C^{gg}_{ii}+C^{gI}_{ii}C^{gI}_{ii}\right]+dC^{GG,N}_{ii}C^{gg}_{ii}+gC^{gG}_{ii}C^{gG}_{ii}+2hC^{gG}_{ii}C^{gI}_{ii}\nonumber\\
&&+\left[c\left(C^{GG}_{ii}+C^{II}_{ii}\right)+eC^{GG,N}_{ii}\right]C^{gg,N}_{ii}\Big)C^{gg,N}_{ii}\Big\}.\label{eq:ester4}
\end{eqnarray}
$C^{gg,N}_{ii}=1/\bar{n}_i$ and $C^{GG,N}_{ii}=\gamma^2_{rms}/\bar{n}_i$, where $\bar{n}_i$ is the average number density of galaxies in the \emph{i}-th redshift bin. Unlike the GI self-calibration, $\Delta B^{Igg}_{iii}$ is dependent on the intrinsic alignment contamination through $C^{Ig}_{ii}$. However, it is still insensitive to the intrinsic alignment contamination in the limit where $C^{GG}_{ii}$ is dominant. 

The errors $\Delta B^{Igg}_{iii}$ and $\Delta C^{Ig}_{ii}$ propagate into the measurement of $B^{IGG}_{iii}$ through Eq. \ref{eq:scale4}. Performing a standard error propagation gives a residual statistical error $\Delta B^{IGG}_{ijk}$. For the equilateral case, this simplifies to
\begin{eqnarray}
\left(\Delta B^{IGG}_{ijk}(\ell)\right)^2&=&\left(\frac{W_{ijk}}{(b^{i}_{1})^2\Pi_{iii}}\right)^2\left(\Delta B^{Igg}_{iii}(\ell)\right)^2+\left(\frac{b^{i}_{2}W_{ijk}}{(b^{i}_{1})^3\Pi_{ii}\omega_{ii}}\right)^2\nonumber\\
&&\times\left[\left(2C^{GG}_{ii}(\ell)\Delta C^{Ig}_{ii}(\ell)\right)^2+\left(2C^{Ig}_{ii}(\ell)C^{GG,N}_{ii}(\ell)\right)^2+\left(\frac{\omega_{ii}}{b^i_1\Pi_{ii}}\right)^2\left(2C^{Ig}_{ii}(\ell)\Delta C^{Ig}_{ii}(\ell)\right)^2\right],\label{eq:ester5}
\end{eqnarray}
where we have neglected terms of order $\Delta^2$. To find the fractional error $\Delta f^{(a)}_{ijk}$ this induces in the lensing bispectrum, we simply scale $\Delta B^{IGG}_{ijk}$ by the factor $f^I_{ijk}$ such that $\Delta f^{(a)}_{ijk}=f^I_{ijk}\Delta B^{IGG}_{ijk}$. Like the 2-point case, this error is equal to $f^{thresh}_{ijk}$, the minimum intrinsic alignment $f^I_{ijk}$ which can be detected through the self-calibration with S/N=1 or $\Delta B^{Igg}_{iii}=B^{Igg}_{iii}$. Thus $f^{thresh}_{ijk}=\Delta f^{(a)}_{ijk}$ represents for the self-calibration both the residual statistical error in the measurement of $C^{GGG}_{ijk}$ and the lower limit at which the intrinsic alignment can be calculated and removed. The GGI self-calibration technique can then turn a systematic contamination $f^I_{ijk}$ of the lensing signal into a statistical error $\Delta f^{(a)}_{ijk}<f^I_{ijk}$ which is insensitive to the original intrinsic alignment contamination.

We compare the error $\Delta f^{(a)}_{ijk}$ to the minimum rms error due to cosmic variance and shot noise in the $B^{GGG}_{ijk}$ measurement, which ignores other sources of error like the intrinsic alignment. The rms error of $B^{GGG}_{ijk}$ ($i\ne j\ne k$) is

\begin{eqnarray}
\left(\Delta B^{GGG}_{ijk}\right)^2&=&\left(C^{GG}_{ii}+C^{GG,N}_{ii}\right)\left(C^{GG}_{jj}+C^{GG,N}_{jj}\right)\left(C^{GG}_{kk}+C^{GG,N}_{kk}\right)+\left(C^{GG}_{ii}+C^{GG,N}_{ii}\right)C^{GG}_{jk}C^{GG}_{jk}\nonumber\\
&&+\left(C^{GG}_{jj}+C^{GG,N}_{jj}\right)C^{GG}_{ik}C^{GG}_{ik}+\left(C^{GG}_{kk}+C^{GG,N}_{kk}\right)C^{GG}_{ij}C^{GG}_{ij}+2C^{GG}_{ij}C^{GG}_{jk}C^{GG}_{ik}.\label{eq:ester7}
\end{eqnarray}
This gives an absolute lower limit on the fractional measurement error of 

\begin{eqnarray}
\left(\epsilon^{min}_{ijk}\right)^{2}=\frac{2\pi^2}{\ell_1\ell_2\ell_3\Delta\ell_1\Delta\ell_2\Delta\ell_3f_{sky}}\left(\frac{\Delta B^{GGG}_{ijk}}{B^{GGG}_{ijk}}\right)^2.\label{eq:ester8}
\end{eqnarray}

Where $\Delta f^{(a)}_{ijk}<\epsilon_{ijk}$, the residual measurement error introduced after the GGI self-calibration is negligible, with very little loss of cosmological information. We find this to be true for an LSST-like survey, as shown in Fig. \ref{fig:merror}. More generally, since $\Delta f^{(a)}_{ijk}$ and $\epsilon_{ijk}$ scale similarly with respect to survey parameters, this should hold for other lensing surveys as well.

\begin{figure}
\includegraphics[angle=270,scale=0.7]{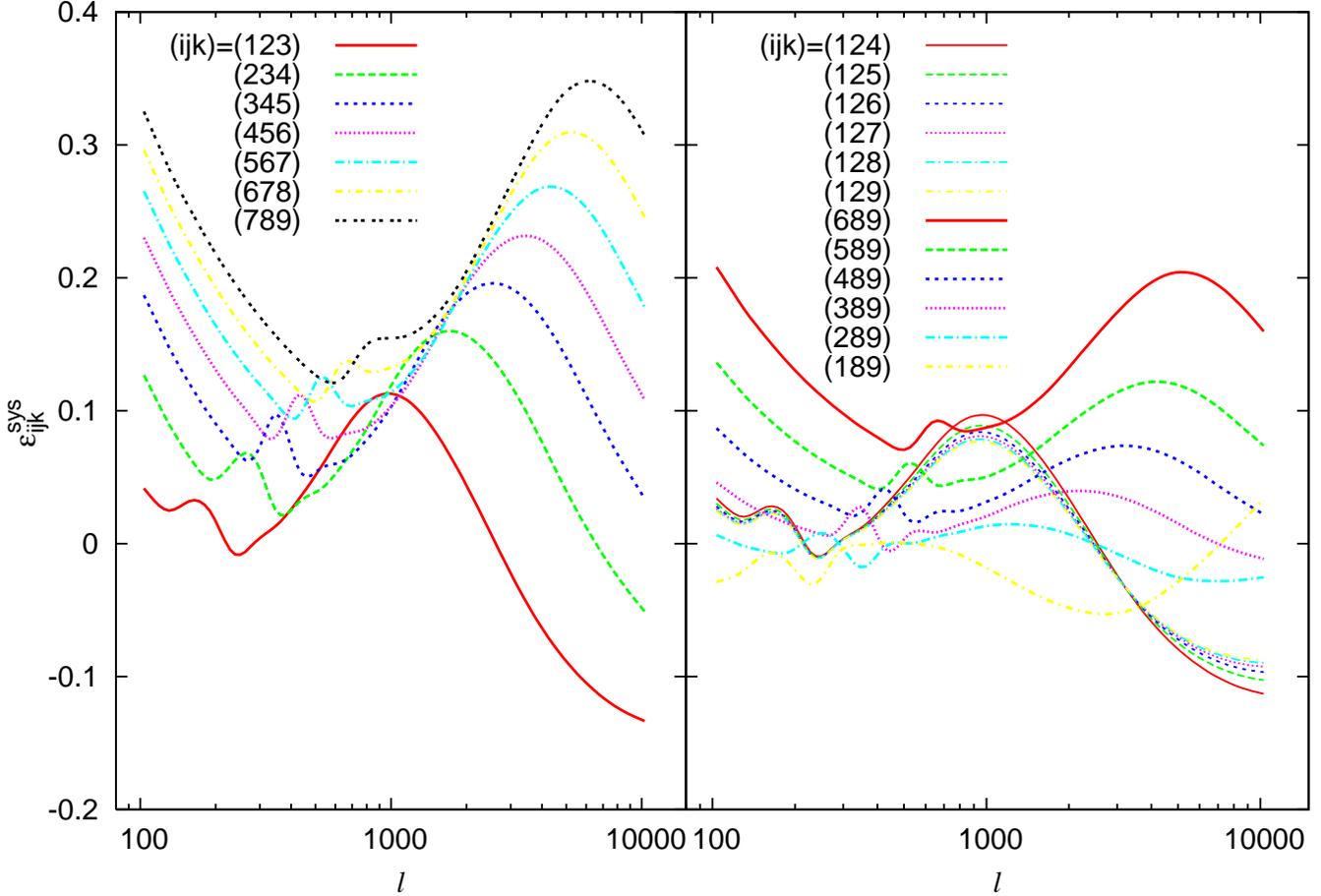}%
\caption{\label{fig:serror}The inaccuracy of the relationship between $B^{IGG}_{ijk}$ and the observable $B^{Igg}_{iii}$ is quantified in Eq. \ref{eq:sys1} by $\epsilon^{sys}_{ijk}$. This inaccuracy is the source of the dominant systematic error in the measurement of $B^{GGG}_{ijk}$ due to the GGI self-calibration technique. Left: $\epsilon^{sys}_{ijk}$ is plotted for three adjacent redshift bins, where the stronger dependence of the lensing kernel on redshift causes a significantly higher inaccuracy. Right: $\epsilon^{sys}_{ijk}$ is plotted for redshift bins of varying distance from each other. As expected, the inaccuracy for these bin choices is generally less than for three adjacent bins. $\epsilon^{sys}_{ijk}$ is plotted for equilateral triangles ($\ell=\ell_1=\ell_2=\ell_3$) in all cases. Equation \ref{eq:scale4} is usually accurate to within $20\%$, except for some adjacent bin choices, where it reaches a maximum of approx. $35\%$. Despite the inaccuracy of Eq. \ref{eq:scale4} being greater than for Eq. \ref{eq:2scale} in the GI self-calibration, the GGI self-calibration is still expected to reduce the GGI intrinsic alignment contamination by a factor of $5$-$10$ or more for all but a few adjacent redshift bin triplets. These results are insensitive to the original intrinsic alignment contamination, such that for any $f^{thresh}_{ijk}<f^I_{ijk}<1$, the GGI self-calibration will reduce the GGI contamination down to survey limits or by a factor of $5$-$10$ or greater, whichever is less, for all but a few adjacent redshift bin triplets.}
\end{figure}
\subsection{The accuracy of the $B^{IGG}_{ijk}$-$B^{Igg}_{iii}$ relation}\label{scalingerror}
In addition to the measurement error introduced through the estimator $\hat{B}^{Igg}_{iii}$, there is a systematic error which is introduced by Eq. \ref{eq:scale4}, which relates the intrinsic alignment contamination $B^{IGG}_{ijk}$ in the lensing bispectrum to other survey observables. The accuracy of Eq. \ref{eq:scale4} is quantified by

\begin{eqnarray}
\epsilon^{sys}_{ijk}&\equiv & \Bigg(\frac{W_{ijk}}{(b^{i}_{1})^2\Pi_{iii}}\frac{B^{Igg}_{iii}(\ell_1,\ell_2,\ell_3)}{B^{IGG}_{ijk}(\ell_1,\ell_2,\ell_3)}-\frac{b^{i}_{2}}{(b^{i}_{1})^2}\frac{W_{ijk}}{\omega_{ii}\Pi_{ii}}\frac{1}{B^{IGG}_{ijk}(\ell_1,\ell_2,\ell_3)}\nonumber\\
&&\times\left[C^{Ig}_{ii}(\ell_1)C^{GG}_{ii}(\ell_2)+C^{GG}_{ii}(\ell_2)C^{Ig}_{ii}(\ell_3)+\frac{\omega_{ii}}{b^i_1\Pi_{ii}}C^{Ig}_{ii}(\ell_1)C^{Ig}_{ii}(\ell_3)\right]\Bigg)^{-1}-1.\label{eq:sys1}
\end{eqnarray}
This induces a residual systematic error in the lensing measurement of

\begin{equation}
\delta f_{ijk}=\epsilon^{sys}_{ijk}f^I_{ijk}.\label{eq:sys2}
\end{equation}

$\epsilon^{sys}_{ijk}$ is evaluated numerically and shown in Fig. \ref{fig:serror} for equilateral triangles. As in the 2-point case, Eq. \ref{eq:scale4} is most accurate for those galaxy triplets which do not share neighbouring redshift bins. In the cases of neighboring bins, the lensing kernel varies more quickly due to the proximity of the galaxies in redshift. This causes Eq. \ref{eq:scale4} to be less accurate, increasing the systematic error. For galaxy triplets with bins which are not adjacent, $|\epsilon^{sys}_{ijk}|<0.1$. For these bin choices, the intrinsic alignment contamination can be suppressed by a factor of 10 or greater. In most cases where two or three bins are adjacent, $|\epsilon^{sys}_{ijk}|<0.2$, which allows for a suppression in the contamination by a factor of 5-10. In only a few of the cases where all three bins are adjacent is $|\epsilon^{sys}_{ijk}|>0.2$, and even in these cases we expect a suppression in the contamination by a factor of 3 or more. These results are insensitive to the original intrinsic alignment contamination, such that for any $f^{thresh}_{ijk}<f^I_{ijk}<1$, the GGI self-calibration will reduce the GGI contamination down to survey limits or by a factor of 5-10 or greater, whichever is less, for all but a few redshift bin triplets.
\subsection{The magnification bias}\label{mag}
In addition to distorting the shapes of galaxies, gravitational lensing introduces a magnification bias to the observed galaxy overdensity $\delta_g^L=\delta_g+2(\alpha-1)\kappa$, where $\alpha$ is determined by the logarithmic slope of the unlensed galaxy luminosity function. The magnification bias affects all three observable bispectra, but we expect the dominant contribution to occur in $B^{(2)}_{iii}$. Including the average magnification bias in the \emph{i}-th redshift bin, $m_i=\langle 2(\alpha-1)\rangle$, Eq. \ref{eq:two} is modified to be

\begin{eqnarray}
B^{(2)}_{iii}=B^{Ggg}_{iii}+B^{Igg}_{iii}+2m_i\left(B^{GGg}_{iii}+B^{IGg}_{iii}\right)+m_i^2\left(B^{GGG}_{iii}+B^{IGG}_{iii}\right).\label{eq:twomag}
\end{eqnarray}

We seek to measure $B^{Igg}_{iii}$ for the GGI self-calibration, so we will examine the effect magnification bias has on this measurement. Applying the estimator in Eq. \ref{eq:qfact}, the bispectra $B^{Igg}_{iii}$ and $B^{GGG}_{iii}$ are unaffected, while the others are suppressed by a factor similar to $(1-Q_3)$. Thus the estimator acts to measure the dominant combination $B^{Igg}_{iii}+m_i^2B^{GGG}_{iii}$, where $m_i^2B^{GGG}_{iii}$ contaminates the $B^{Igg}_{iii}$ measurement. Because the GGI self-calibration depends on the results of the 2-point self-calibration, we will also require the contribution $C^{Ig}_{ii}+m_iC^{GG}_{ii}$ from $C^{(2)}_{ii}$ as discussed by \cite{22}.
 
We cannot remove this contamination with any certainty due to measurement errors on $m_i$, $C^{GG}_{ii}$ and $B^{GGG}_{iii}$. The direct estimation of the errors involved is lengthy, so we will instead determine the accuracy to which these measurements must be made in order for the contribution due to magnification bias to be negligible with respect to other errors in the GGI self-calibration.

We will assume $m_i$ has some measurement error $\Delta m_i$, $B^{GGG}_{iii}$ a measurement error $\Delta B^{GGG}_{iii}$ and $C^{GG}_{ii}$ a measurement error $\Delta C^{GG}_{ii}$. From Eqs. \ref{eq:scale4} \& \ref{eq:2scale}, the induced measurement error in $B^{IGG}_{ijk}$ is

\begin{eqnarray}
\Delta B^{IGG}_{ijk}&=&\frac{W_{ijk}}{(b^{i}_{1})^2\Pi_{iii}}\left(2m_i\Delta m_i B^{GGG}_{iii}+m_i^2B^{GGG}_{iii}\Delta B^{GGG}_{iii}\right)-\frac{b^{i}_{2}W_{ijk}}{(b^{i}_{1})^3\Pi^2_{ii}}\left(2\frac{W_{ij}}{\omega_{ii}}+1\right)\left(\Delta m_iC^{GG}_{ii}+m_iC^{GG}_{ii}\Delta C^{GG}_{ii}\right)^2.\label{eq:mag1}
\end{eqnarray}
Since we are only interested in the upper limit of this effect, we note that $B^{GGG}_{ijk}>B^{GGG}_{iii}$ for $i<j<k$ and use the reduced bispectrum for equilateral triangles $B^{GGG}_{iii}\equiv 3Q_kC^{GG}_{ii}C^{GG}_{ii}$ to write a simplified expression for the induced fractional error in the $B^{GGG}_{ijk}$ measurement as

\begin{eqnarray}
\Delta f^M_{ijk}&<&\frac{W_{ijk}}{(b^{i}_{1})^2\Pi_{iii}}\left(2\left|m_i\Delta m_i \right|+m_i^2\left|\frac{\Delta B^{GGG}_{iii}}{B^{GGG}_{iii}}\right|\right)-\frac{b^{i}_{2}W_{ijk}}{3Q_k(b^{i}_{1})^3\Pi^2_{ii}}\left(2\frac{W_{ij}}{\omega_{ii}}+1\right)\left(\left|\Delta m_i\right|+\left|m_i\frac{\Delta C^{GG}_{ii}}{C^{GG}_{ii}}\right|\right)^2\nonumber\\
&<&O(10^{-4})\left[\left(2\left|m_i\frac{\Delta m_i}{0.1} \right|+m_i^2\left|\frac{\Delta B^{GGG}_{iii}/B^{GGG}_{iii}}{10\%}\right|\right)+\left(\left|\frac{\Delta m_i}{0.1}\right|+\left|m_i\frac{\Delta C^{GG}_{ii}/C^{GG}_{ii}}{10\%}\right|\right)^2\right].\label{eq:mag2}
\end{eqnarray}

The above expression is an upper limit on the magnitude of $\Delta f^M_{ijk}$ given any choice of $i,j,k$. For a $m_i$ which is large enough to be non-negligible, we need only require an accuracy in its measurement of $\Delta m_i=0.1$ and a measurement accuracy for $B^{GGG}_{iii}$ and $C^{GG}_{iii}$ of 10\% in order to have $f^M_{ijk}<O(10^{-4})$, which is safely negligible by a factor of 10 compared to the minimum measurement error $\epsilon^{min}_{ijk}$ of the lensing bispectrum. As discussed by Zhang, this level of accuracy can likely be accomplished by direct measurement of $m_i$ under the approximation $C^{(1)}_{ii}\approx C^{GG}_{ii}$ (see Eq. \ref{eq:2obs}) if the lensing contamination $C^{II}_{ii}<10\%$. However, if the II contamination is greater than $10\%$ of the lensing signal, more detailed methods must be employed to achieve a great enough accuracy in the $m_i$ measurement for it to be safely negligible, some of which are discussed by \cite{22}.

\subsection{Non-Gaussianity and galaxy bias }\label{bias}

We are only interested in the bispectrum due to the non-linear evolution of gravitational clustering and the associated intrinsic alignment contamination, leaving the accurate estimation of the bispectrum due to primordial non-Gaussianity to other works. Equation \ref{eq:onea} should then include a term $B^{NG_0}_{ijk}$ which must be separately accounted for. Similarly, the relation between the 3D matter bispectrum and 3D galaxy bispectrum depends on non-Gaussianity beyond the scale dependent correction $b_1(z)\rightarrow b_1(z)+\Delta b(k,z)$ used in relating the 3D matter power spectrum to the 3D galaxy power spectrum \citep{28}. Equation \ref{eq:bg} must also include the contributions by non-Gaussianity in the term $B^{NG_0}_{ijk}$ and from the trispectrum which we have previously neglected. 

\begin{figure}
\includegraphics[angle=270,scale=0.65]{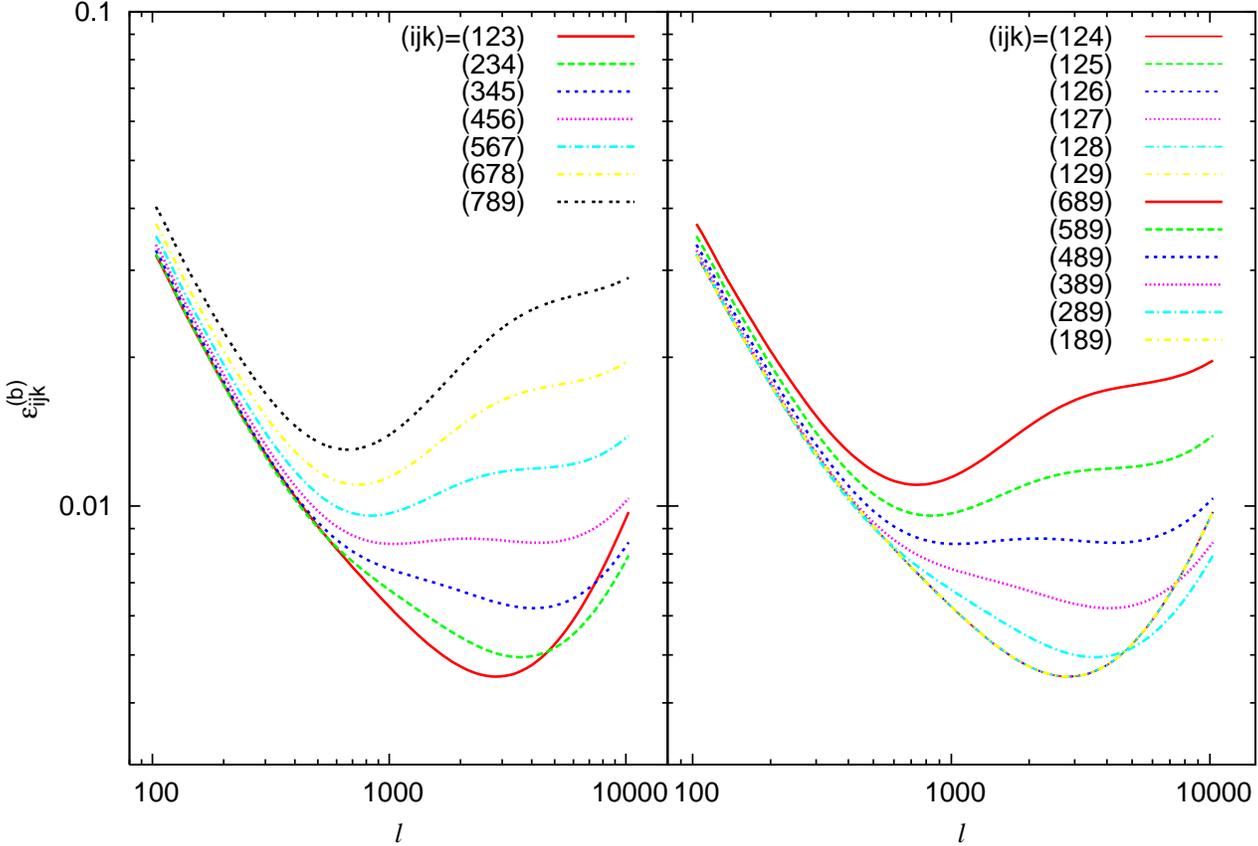}%
\caption{\label{fig:bias}The fractional error $\epsilon^{(b)}_{ijk}=\Delta B^{IGG}/B^{IGG}$ in Eq. \ref{eq:scale4} is shown for equilateral triangles ($\ell=\ell_1=\ell_2=\ell_3$) due to uncertainties in the linear and non-linear galaxy bias. $\epsilon^{(b)}_{ijk}$ is typically less than 2\% except for large scales. The measurement error this induces in the final measurement of $B^{GGG}$, $\Delta f^{(b)}_{ijk}=\epsilon^{(b)}_{ijk}f^I_{ijk}$, is reduced by the factor $f^I_{ijk}\le 1$. It is thus typically negligible when compared to the minimum measurement error $\epsilon^{min}_{ijk}$ in $B^{GGG}$ (Fig. \ref{fig:merror}).}
\end{figure}

The full expression including all non-Gaussian contributions is given in Appendix B of \cite{28}. However, from Figs. 10-14 of \cite{28}, it is clear that if we avoid very stretched or elongated triangle shapes that are very sensitive to non-Gaussianity, at the scales of interest in a lensing survey ($10^2<\ell<10^4$), the total contribution to the relation by non-Guassianity as a fraction of the non-linear term is less than 10\% for $f_{NL}=40$ and $g_{NL}=10^4$. If we accept the smaller values of $f_{NL}=4$ and $g_{NL}=100$, this fractional contribution is less than 1\%. Thus we can safely ignore the contribution of the non-Gaussianity as a source of error to the relation since it is expected to be on the order of the minimum GG measurement error and less than the systematic error discussed in Sec. \ref{scalingerror}. Future work will better constrain and model the effects of non-Gaussianity, thus allowing its effect to be fully accounted for in the GGI self-calibration.

The linear galaxy bias is discussed by \cite{22}, and the error induced by the expected uncertainty in its measurement in the GI self-calibration is demonstrated to be negligible compared to other sources of error. The linear and non-linear galaxy bias terms can be measured simultaneously by using the approach of \cite{32}. Using measurements of $C^{gg}$ and $B^{ggg}$, we extract the bias information from the relationship

\begin{equation}
B_{iii}^{ggg}(\ell_1,\ell_2,\ell_3)\approx(b^i_1)^3B^{mmm}_{iii}+(b^i_1)^2b^i_2\left(C^{mm}_{ii}(\ell_1)C^{mm}_{ii}(\ell_2)+C^{mm}_{ii}(\ell_2)C^{mm}_{ii}(\ell_3)+C^{mm}_{ii}(\ell_1)C^{mm}_{ii}(\ell_3)\right),\label{eq:bias}
\end{equation}
where $C^{mm}$ and $B^{mmm}$ are the angular matter power and bispectrum, weighted identically to galaxies. Equation \ref{eq:bias} is the analog to Eq. \ref{eq:bg}, which includes an intrinsic alignment component. The matter power and bispectrum can be tightly constrained by CMB measurements and then evolved, given a cosmology, to low redshift to predict $C^{mm}$ and $B^{mmm}$.

Both measurement error in $B_{iii}^{ggg}$ and uncertainties in the predictions of $C^{mm}$ and $B^{mmm}$ will affect the measurement of the linear and non-linear galaxy bias parameters. The non-linear galaxy bias is more difficult to constrain precisely than the linear galaxy bias, with typical measured and expected uncertainties in its measurement of up to $\Delta [b^i_2/(b^i_2)^2]\approx 0.5$ at $1\sigma$ confidence \citep{29,31}. We use the estimate of \cite{22} for the measurement error in the linear galaxy bias 
\begin{equation}
\frac{\Delta b_1^i}{b_1^i}\approx \frac{1}{2}\sqrt{\frac{1}{\ell\Delta\ell f_{sky}}}\left(1+\frac{C^{gg,N}}{C^{gg}}\right)
\end{equation}
to plot in Fig. \ref{fig:bias} the fractional error $\epsilon^{(b)}_{ijk}=\Delta B^{IGG}/B^{IGG}$ in Eq. \ref{eq:scale4} of even a large uncertainty for the non-linear galaxy bias of $\Delta b^i_2\approx 0.5$. We find that $\epsilon^{(b)}_{ijk}$ is generally less than 2\% except for large scales. The measurement error this induces in the final measurement of $B^{GGG}$ is then $\Delta f^{(b)}_{ijk}=\epsilon^{(b)}_{ijk}f^I_{ijk}$. Even for very large $f^I_{ijk}=1$, $\Delta f^{(b)}_{ijk}$ is typically comparable to or less than the minimum measurement error $\epsilon^{min}_{ijk}$ in $B^{GGG}$ (Fig. \ref{fig:merror}). For a typical $f^I_{ijk}$, we would expect it to be entirely negligible for all bin choices.

The only real limitation which comes from the galaxy bias is then the scale to which it can be applied in the non-linear regime. Recent work \citep{31} has shown that the scale down to which the bias model we have employed is accurate can be extended to $k=0.5$, which corresponds to $\ell\approx 1000$ at the median redshift of an LSST-like survey. Future work may extend this range further, but for now this places an approximate upper limit on the $\ell$ at which the self-calibration can function to a high degree of accuracy. In the future, a more robust bias model could be chosen for the very highly non-linear regime to extend this limit with relative ease, as it will alter only the form of Eq. \ref{eq:scale4} and the resulting performance calculations, while the method of extracting $B^{Igg}_{iii}$ remains unchanged.

\subsection{Other sources of uncertainty}\label{other}

The GGI self-calibration requires the calculation of $W_{ijk}$ and $Q_3$, which include the cosmology-dependent lensing kernel. This introduces an uncertainty due to the measurement of $\Omega_m$ and the distance-redshift relation. However, we expect this uncertainty to be negligible when compared to other dominant sources of error in the GGI self-calibration. \cite{33} have measured $\Omega_m$ to 5\% accuracy, and new measurements are expected to constrain $\Omega_m$ to 1-2\%. The distance-redshift relation will also be constrained to 1\% by baryon acoustic oscillations and supernovae \citep{34}. We expect that given these constraints, any uncertainty introduced by the lensing kernel will only affect the GGI self-calibration at the percent level, which is negligible compared to the expected systematic error $\delta f_{ijk}$ of Eq. \ref{eq:sys2}. An iterative approach can also be applied, where a set of initial cosmological parameters is chosen as above and used for the 2- and 3-point self-calibration, from which new (improved) parameter constraints can be calculated and applied again until the interactive process converges.

Similarly, we have used an approximate fitting formula derived from perturbation theory by \cite{30} for the bispectrum in our error estimations. This is only expected to be accurate to within 15\% when compared to N-body simulations for the lensing bispectrum. We thus expect uncertainty due to the calculation of the bispectrum to be dominant when compared to errors associated with the power spectrum calculation. A more accurate approach to modelling the bispectrum and the effects of intrinsic alignment will provide more accurate estimates of the GGI self-calibration performance, which we leave to a later work.

Catastrophic photo-z error also affects the GGI self-calibration through the assumed galaxy distribution. We assume a Gaussian photo-z PDF in our numerical calculations, but observed photo-z PDFs generally have non-negligible outliers. This affects the GGI self-calibration through the calculation of $Q_3$ and the relationship between $B^{IGG}$ and $B^{Igg}$. However, these effects are suppressed due to both numerator and denominator being affected in similar ways. The effect can be further decreased by better photo-z PDF template estimates and better calibration of photo-z errors, and we expect the GGI self-calibration to ultimately be safe from non-negligible degradation due to catastrophic photo-z errors.

The relationship between $B^{IGG}$ and $B^{Igg}$ depends upon our assumption of a deterministic galaxy bias, which is not perfectly accurate in real galaxy distributions. This could cause both random and systematic error in the GGI self-calibration. A true quantification of this effect is beyond the scope of this paper, as the possible correlation between stochasticity and intrinsic alignment is not well understood. However, \cite{35} has shown that it is possible to suppress the galaxy stochasticity to the 1\% level in some cases, which allows that the effect of stochasticity in the GGI self-calibration could ultimately be limited to the percent level, which would be safely negligible compared to other sources of error.
\subsection{Summary of residual errors}\label{behavior}
There are three regimes under which the performance of the GGI self-calibration can be summarised. These are defined by the magnitude of the GGI contamination as represented by $f^I_{ijk}$. The first is where the ggI correlation is too small to detect in $B^{(2)}$, with $f^I_{ijk}\le f^{thresh}_{ijk}$. If the intrinsic alignment cannot be detected in $B^{(2)}$, the GGI self-calibration is not applicable. This generally means that the GGI contamination is also negligible when compared to $\epsilon^{min}_{ijk}$, the minimum statistical error in the lensing bispectrum, and there is no need to correct for it.

If $f^I_{ijk}>f^{thresh}_{ijk}$, then the GGI contamination to the lensing bispectrum is likely not negligible, and it must be corrected for. The GGI self-calibration is now able to detect and calculate the GGI correlation. In the second regime, where $\Delta f^{thresh}_{ijk}>\epsilon^{sys}_{ijk}f^I_{ijk}$, the statistical error $\Delta f^{(a)}_{ijk}$ induced by measurement error in the estimator $\hat{B}^{Igg}_{iii}$ is dominant. As shown in Fig. \ref{fig:merror}, this error is generally negligible when compared to $\epsilon^{min}_{ijk}$, and so in this regime, the GGI self-calibration should perform at the statistical limit of the lensing survey.

Finally, where $\Delta f^{thresh}_{ijk}<\epsilon^{sys}_{ijk}f^I_{ijk}$, the systematic error $\delta f_{ijk}=\epsilon^{sys}_{ijk}f^I_{ijk}$ due to the relationship between $B^{IGG}_{ijk}$ and $B^{Igg}_{iii}$ in Eq. \ref{eq:scale4} is dominant. In the case where $\epsilon^{sys}_{ijk}<\epsilon^{min}_{ijk}/f^I_{ijk}$, $\epsilon^{min}_{ijk}$ is still dominant. Otherwise the GGI self-calibration can suppress the GGI contamination by a factor of 5-10 or more for all but a few adjacent redshift bin choices. In this case, other complementary techniques could be employed to further reduce the GGI contamination down to the statistical limit for the lensing survey.

In the 2-point correlations, one such case has been explored by \cite{23}, but such studies of the 3-point intrinsic alignment are left to be done. \cite{36} combines the GI self-calibration with a photo-z self-calibration to better protect the GI self-calibration against catastrophic photo-z effects. Both methods are possible because the GI and GGI self-calibration uses primarily those correlations in one redshift bin to estimate the intrinsic alignment, while \cite{23,36} use those correlations between redshift bins. As first mentioned in Sec. \ref{intro}, others have also used information between redshift bins to calibrate the intrinsic alignment contamination in the 2- and 3-point correlations \citep{14b,19,20a,20b,21,9b}. Such techniques for the 3-point intrinsic alignment correlations should eventually complement the GGI self-calibration for improved reductions in the contamination by the intrinsic alignment in the cosmic shear signal, but much work is left to be done.
\section{Conclusion}\label{conc}
The GGG bispectrum has been shown to be strongly contaminated by the 3-point intrinsic alignment correlations. While the III and GII correlations can be neglected by considering only the cross-correlation bispectrum between three different redshift bins, the GGI correlation remains a contaminant. \cite{22} first proposed the self-calibration technique in order to calculate and remove the 2-point GI contamination from the GG power spectrum. In this work we verify the performance of the GI self-calibration technique, and expand the self-calibration to the 3-point correlations, proposing the GGI self-calibration technique to calculate and remove the GGI correlation from the GGG bispectrum. 

We first establish the estimator $\hat{B}^{Igg}_{iii}$ to extract the ggI correlation from the galaxy ellipticity-density-density measurement for a photo-z galaxy sample. We show that this estimator is expected to be generally applicable to weak lensing surveys and reduces to the simple extraction method for spectroscopic galaxy samples at low photo-z error. We then develop a relation between the GGI and ggI bispectra using the linear and non-linear galaxy bias to relate the galaxy density and cosmic shear measurements. This allows us to calculate and remove the GGI correlation from the GGG bispectrum. While this method is in principle applicable to all $\ell$ and triangle shapes, we do note some modest restrictions in section \ref{bias} on very elongated triangles due to the effects of non-Gaussianity and at very non-linear scales due to limitations in the understanding of the galaxy bias model used.

We quantify the performance of the GGI self-calibration technique for a typical weak-lensing survey, using anticipated parameters for the LSST as an example case. The residual statistical error due to measurement uncertainty in the estimator $\hat{B}^{Igg}_{iii}$ is shown to be generally negligible when compared to the minimum measurement error in the lensing bispectrum. By considering the systematic error introduced by the relationship between $B^{IGG}_{ijk}$ and $B^{Igg}_{iii}$, we show that for galaxy triplets with bins which are not adjacent, $|\epsilon_{ijk}|<0.1$. For these bin choices, the intrinsic alignment contamination can be suppressed by a factor of 10 or greater. In most cases where two or three bins are adjacent, $|\epsilon_{ijk}|<0.2$, which allows for a suppression in the contamination by a factor of 5. In only a few of the cases where all three bins are adjacent is $|\epsilon_{ijk}|>0.2$, and even in these cases we expect a suppression in the contamination by a factor of 3 or more. This will potentially allow the GGI self-calibration to reduce the GGI correlation to the statistical limit of the lensing survey, as discussed in Sec. \ref{behavior}. 

These results are insensitive to the original intrinsic alignment contamination, such that for any $f^{thresh}_{ijk}<f^I_{ijk}<1$, the GGI self-calibration will reduce the GGI contamination down to survey limits or by a factor of 5-10 or greater, whichever is less, for all but a few adjacent redshift bin triplets. This is only slightly reduced from the GI self-calibration, where for any $f^{thresh}_{ij}<f^I_{ij}<1$, the GI self-calibration reduces the GI contamination down to survey limits or by a factor of 10 or greater, whichever is less. We thus expect the GGI self-calibration to perform near the level of the GI self-calibration, and together they promise to be an efficient technique to isolate both the 2- and 3-point intrinsic alignment signals from the cosmic shear signal.

\section*{Acknowledgments}
We thank E. Komatsu, R. Mandelbaum, and P. Zhang for useful comments. MI acknowledges that this material is based upon work supported in part by National Science Foundation under grant AST-1109667 and NASA under grant NNX09AJ55G, and that part of the calculations for this work have been performed on the Cosmology Computer Cluster funded by the Hoblitzelle Foundation.
\appendix
\section{Calculation of coefficients in $\Delta B^{Igg}_{\MakeLowercase{iii}}$}

Upon evaluating the sum and taking the Fourier transform of Eq. \ref{eq:ester2}, each of the products of the correlations have a numerical coefficient due to the restrictions on redshift ordering. Many, however, are identical due to symmetries. The calculation of the unique coefficients \emph{a-h} in Eqs. \ref{eq:ester3} \& \ref{eq:ester4} are summarised here. The first coefficient is trivial, due to products with no noise correlations or correlations like $\langle\delta_{\alpha}\kappa_{\nu}\rangle$, which are themselves orientation dependent. We then calculate for a term like $\langle\delta_{\alpha}\delta_{\lambda}\rangle\langle\delta_{\beta}\delta_{\mu}\rangle\langle\kappa_{\gamma}\kappa_{\nu}\rangle$

\begin{eqnarray}
\frac{N_P^{-6}}{(1-Q_3)^2}\sum_{\alpha\beta\gamma}\sum_{\lambda\mu\nu}(3S_{\alpha\beta\gamma}-Q_3)(3S_{\lambda\mu\nu}-Q_3)\approx 1.
\end{eqnarray}
For terms like $\langle\delta^N_{\alpha}\delta^N_{\lambda}\rangle\langle\delta_{\beta}\delta_{\mu}\rangle\langle\kappa_{\gamma}\kappa_{\nu}\rangle$ $\propto\delta_{\alpha\lambda}$, which include one galaxy density noise correlation

\begin{eqnarray}
a\equiv\frac{N_P^{-5}}{(1-Q_3)^2}\sum_{\alpha\beta\gamma}\sum_{\mu\nu}(3S_{\alpha\beta\gamma}-Q_3)(3S_{\alpha\mu\nu}-Q_3)\approx 1+\frac{1}{5(1-Q_3)^2}.
\end{eqnarray}
For terms like $\langle\delta_{\alpha}\delta_{\lambda}\rangle\langle\delta_{\beta}\delta_{\mu}\rangle\langle\kappa^N_{\gamma}\kappa^N_{\nu}\rangle$ $\propto\delta_{\gamma\nu}$, which include one convergence noise correlation

\begin{eqnarray}
b\equiv\frac{N_P^{-5}}{(1-Q_3)^2}\sum_{\alpha\beta\gamma}\sum_{\lambda\mu}(3S_{\alpha\beta\gamma}-Q_3)(3S_{\lambda\mu\gamma}-Q_3)\approx 1+\frac{4}{5(1-Q_3)^2}.
\end{eqnarray}
For terms like $\langle\delta^N_{\alpha}\delta^N_{\lambda}\rangle\langle\delta^N_{\beta}\delta^N_{\mu}\rangle\langle\kappa_{\gamma}\kappa_{\nu}\rangle$ $\propto\delta_{\alpha\lambda}\delta_{\beta\mu}$, which include two galaxy density noise correlations

\begin{eqnarray}
c\equiv\frac{N_P^{-4}}{(1-Q_3)^2}\sum_{\alpha\beta\gamma}\sum_{\nu}(3S_{\alpha\beta\gamma}-Q_3)(3S_{\alpha\beta\gamma}-Q_3)\approx 1-\frac{1}{4(1-Q_3)^2}.
\end{eqnarray}
For terms like $\langle\delta^N_{\alpha}\delta^N_{\lambda}\rangle\langle\delta_{\beta}\delta_{\mu}\rangle\langle\kappa^N_{\gamma}\kappa^N_{\nu}\rangle$ $\propto\delta_{\alpha\lambda}\delta_{\gamma\nu}$, which include one galaxy density noise correlation and one convergence noise correlation

\begin{eqnarray}
d\equiv\frac{N_P^{-4}}{(1-Q_3)^2}\sum_{\alpha\beta\gamma}\sum_{\mu}(3S_{\alpha\beta\gamma}-Q_3)(3S_{\alpha\mu\gamma}-Q_3)\approx 1+\frac{5}{4(1-Q_3)^2}.
\end{eqnarray}
For terms like $\langle\delta^N_{\alpha}\delta^N_{\lambda}\rangle\langle\delta^N_{\beta}\delta^N_{\mu}\rangle\langle\kappa^N_{\gamma}\kappa^N_{\nu}\rangle$ $\propto\delta_{\alpha\lambda}\delta_{\beta\mu}\delta_{\gamma\nu}$, which include only noise correlations

\begin{eqnarray}
e\equiv\frac{N_P^{-3}}{(1-Q_3)^2}\sum_{\alpha\beta\gamma}(3S_{\alpha\beta\gamma}-Q_3)^2\approx 1+\frac{2}{(1-Q_3)^2}.
\end{eqnarray}
For terms like $\langle\delta_{\alpha}\delta_{\lambda}\rangle\langle\kappa_{\gamma}\delta_{\mu}\rangle\langle\kappa_{\nu}\delta_{\beta}\rangle$, which include two correlations with the orientation dependence described in Eq. \ref{eq:sub}

\begin{eqnarray}
f&\equiv&\frac{N_P^{-6}}{(1-Q_3)^2}\sum_{\alpha\beta\gamma}\sum_{\lambda\mu\nu}(3S_{\alpha\beta\gamma}-Q_3)(3S_{\lambda\mu\nu}-Q_3)\frac{1}{2} \left(\frac{S_{\mu\gamma}}{(1-Q_2)}+\frac{S_{\gamma\mu}}{Q_2}\right)\frac{1}{2} \left(\frac{S_{\beta\nu}}{(1-Q_2)}+\frac{S_{\nu\beta}}{Q_2}\right)\nonumber\\
&\approx&\frac{5Q_3^2-Q_3(3+13Q_2+2Q_2^2)+5Q_2(1+2Q_2)}{80Q_2^2(1-Q_2)^2(1-Q_3)^2}.
\end{eqnarray}
For terms like $\langle\delta^N_{\alpha}\delta^N_{\lambda}\rangle\langle\kappa_{\gamma}\delta_{\mu}\rangle\langle\kappa_{\nu}\delta_{\beta}\rangle$ $\propto\delta_{\alpha\lambda}$, which include two correlations with the orientation dependence described in Eq. \ref{eq:sub} and one galaxy density noise correlation

\begin{eqnarray}
g&\equiv&\frac{N_P^{-5}}{(1-Q_3)^2}\sum_{\alpha\beta\gamma}\sum_{\mu\nu}(3S_{\alpha\beta\gamma}-Q_3)(3S_{\alpha\mu\nu}-Q_3)\frac{1}{2} \left(\frac{S_{\mu\gamma}}{(1-Q_2)}+\frac{S_{\gamma\mu}}{Q_2}\right)\frac{1}{2} \left(\frac{S_{\beta\nu}}{(1-Q_2)}+\frac{S_{\nu\beta}}{Q_2}\right)\nonumber\\
&\approx&\frac{3+5Q_3^2-Q_3(3+13Q_2+2Q_2^2)+15Q_2^2}{80Q_2^2(1-Q_2)^2(1-Q_3)^2}.
\end{eqnarray}
Finally, for terms like $\langle\delta^N_{\alpha}\delta^N_{\lambda}\rangle\langle\kappa_{\gamma}\delta_{\mu}\rangle\langle I_{\nu}\delta_{\beta}\rangle$ $\propto\delta_{\alpha\lambda}$, which include one correlation with the orientation dependence described in Eq. \ref{eq:sub} and one galaxy density noise correlation

\begin{eqnarray}
h&\equiv&\frac{N_P^{-5}}{(1-Q_3)^2}\sum_{\alpha\beta\gamma}\sum_{\mu\nu}(3S_{\alpha\beta\gamma}-Q_3)(3S_{\alpha\mu\nu}-Q_3)\frac{1}{2} \left(\frac{S_{\mu\gamma}}{(1-Q_2)}+\frac{S_{\gamma\mu}}{Q_2}\right)\nonumber\\
&\approx&\frac{20Q_3^2-5Q_3(5+6Q_2)+6(1+6Q_2)}{80Q_2(1-Q_2)(1-Q_3)^2}.
\end{eqnarray}

For typical values $Q_2=1/2$ and $Q_3=2/5$, these coefficients are

\begin{eqnarray}
a&=&\frac{14}{9}\approx 1.6\nonumber\\
b&=&\frac{29}{9}\approx 3.2\nonumber\\
c&=&\frac{11}{36}\approx 0.3\nonumber\\
d&=&\frac{161}{36}\approx 4.5\nonumber\\
e&=&\frac{59}{9}\approx 6.6\nonumber\\
f&=&1\nonumber\\
g&=&\frac{71}{36}\approx 2.0\nonumber\\
h&=&\frac{14}{9}\approx 1.5.
\end{eqnarray}

\label{lastpage}


\begin{thebibliography}{99}
%
\bibitem[Acquaviva et~al. (2008)]{2a} Acquaviva V., Hajian A., Spergel D., Das S., 2008, PRD, 78, 043514 
\bibitem[Albrecht et~al. (2006)]{34} Albrecht A., et~al., 2006, Report of the Dark Energy Task Force, arXiv:astro-ph/0609591
\bibitem[Bacon, Refregier \& Ellis (2000)]{1a} Bacon D.J., Refregier A.R., Ellis R.S., 2000, MNRAS, 318, 625 
\bibitem[Bacon et~al. (2001)]{6a} Bacon D.J., Refregier A., Clowe D., Ellis R.S., 2001, MNRAS, 325, 1065 
\bibitem[Baldauf et~al. (2010)]{35} Baldauf T., Smith R., Seljak U., Mandelbaum R., 2010, PRD, 81, 063531
\bibitem[Bean \& Tangmatitham (2010)]{2b} Bean R., Tangmatitham M., 2010, PRD, 81, 083534 
\bibitem[Bernstein (2009)]{25} Bernstein G.M., 2009, ApJ, 695, 652
\bibitem[Bernstein \& Jarvis (2002)]{6b} Bernstein G.M., Jarvis M., 2002, AJ, 123, 583 
\bibitem[Blazek, McQuinn \& Seljak (2011)]{8a} Blazek J., McQuinn M., Seljak U., 2011, JCAP, 05, 010
\bibitem[Bridle \& King (2007)]{9a} Bridle S., King L., 2007, New J. Phys. 9, 444
\bibitem[Brown et~al. (2002)]{7} Brown M.L., Taylor A.N., Hambly N.C., Dye S., 2002, MNRAS, 333, 501 
\bibitem[Brown et~al. (2003)]{1b} Brown M.L., Taylor A.N., Bacon D.J., Gray M.E., Dye S., Meisenheimer K., Wolf C., 2003, MNRAS, 341, 100 
\bibitem[Capozziello, Cardone \& Troisi (2006)]{2c} Capozziello S., Cardone V.F., Troisi A., 2006, PRD, 73, 104019
\bibitem[Catelan, Kamionkowski \& Blandford (2001)]{6c} Catelan P., Kamionkowski M., Blandford R.D., 2001, MNRAS, 320, L7 
\bibitem[Cooray \& Sheth (2002)]{27} Cooray A., Sheth R., 2002, Phys. Rept. 372, 1
\bibitem[Crittenden et~al. (2001)]{8c} Crittenden R.G., Natarajan P., Pen U.-L., Theuns T., 2001, ApJ, 559, 552
\bibitem[Croft \& Metzler (2000)]{6d} Croft R., Metzler C., 2000, ApJ, 545, 561
\bibitem[Daniel et~al. (2008)]{2d} Daniel S., Caldwell R., Cooray A., Melchiorri A., 2008, PRD, 77, 103513
\bibitem[Daniel et~al. (2010)]{2e} Daniel S., Linder E., Smith T., Caldwell R., Corray A., Leauthaud A., Lombriser L., 2010, PRD, 80, 123508
\bibitem[Dossett, Moldenhauer \& Ishak (2011)]{2f} Dossett J., Moldenhauer J., Ishak M., 2011, PRD, 84, 023012
\bibitem[Eisenstein, Hu \& Tegmark (1999)]{1c} Eisenstein D.J., Hu W., Tegmark M., 1999, ApJ, 518, 2 
\bibitem[Erben et~al. (2001)]{6e} Erben T., Van Waerbeke L., Bertin E., Mellier Y., Schneider P., 2001, A\&A, 366, 717 
\bibitem[Faltenbacher et~al. (2009)]{14a} Faltenbacher A., Li C., White S.D.M., Jing Y.P., Mao S., Wang J., 2009, Res. Astron. \& Astrophys., 9, 41
\bibitem[Fry (1994)]{32} Fry J., 1994, PRL, 73, 215
\bibitem[Fry \& Gaztanaga (1993)]{26} Fry J.N., Gaztanaga E., 1993, ApJ, 413, 447
\bibitem[Fu, Wu \& Yu (2009)]{1d} Fu X., Wu P., Yu H., 2009, PLB, 677, 12 
\bibitem[Heavens, Refregier \& Heymans (2000)]{6f} Heavens A., Refregier A., Heymans C., 2000, MNRAS, 319, 649 
\bibitem[Heymans \& Heavens (2003)]{8e} Heymans C., Heavens A., 2003, MNRAS, 339, 711
\bibitem[Heymans et~al. (2004)]{6g} Heymans C., Brown M., Heavens A., Meisenheimer K., Taylor A., Wolf C., 2004, MNRAS, 347, 895 
\bibitem[Heymans et~al. (2006)]{13} Heymans C., White M., Heavens A., Vale C., Van Waerbeke L., 2006, MNRAS, 371, 750
\bibitem[Hirata \& Seljak (2003a)]{6h} Hirata C.M., Seljak U., 2003, MNRAS, 343, 459 
\bibitem[Hirata \& Seljak (2003b)]{8f} Hirata C.M., Seljak U., 2003, PRD, 67, 43001
\bibitem[Hirata \& Seljak (2004)]{11} Hirata C.M., Seljak U., 2004, PRD, 70, 063526
\bibitem[Hirata et~al. (2007)]{10} Hirata C.M., Mandelbaum R., Ishak M., Seljak U., Nichol R., Pimbblet K.A., Ross N.P., Wake D., 2007, MNRAS, 381, 1197
\bibitem[Hoekstra et~al. (2002)]{1e} Hoekstra H., Yee H.K.C., Gladders M.D., Barrientos L.F., Hall P.B., Infante L., 2002, ApJ, 72, 55 
\bibitem[Hu (2002)]{1g} Hu W., 2002, PRD, 65, 023003 
\bibitem[Hu \& Tegmark (1999)]{1f} Hu W., Tegmark M., 1999, ApJL, 514, L65 
\bibitem[Huterer \& Linder (2007)]{2g} Huterer D., Linder E., 2007, PRD, 75, 023519
\bibitem[Ishak \& Dossett (2009)]{2i} Ishak M., Dossett J., 2009, PRD, 80, 043004 
\bibitem[Ishak et~al. (2004)]{6i} Ishak M., Hirata C.M., McDonald P., Seljak U., 2004, PRD, 69, 083514
\bibitem[Ishak, Upadhye \& Spergel (2006)]{2h} Ishak M., Upadhye A., Spergel D., 2006, PRD, 74, 043513 
\bibitem[Jarvis et~al. (2003)]{1h} Jarvis M., Bernstein G.M., Fischer P., Smith D., Jain B., Tyson J.A., Wittman D., 2003, AJ, 125, 1014 
\bibitem[Jing (2002)]{8g} Jing Y.P., 2002, MNRAS, 335, 89
\bibitem[Jeong \& Komatsu (2009)]{28} Jeong D., Komatsu E., 2009, ApJ, 703, 1230
\bibitem[Joachimi \& Bridle (2010)]{9b} Joachimi B., Bridle S., 2010, A\&A, 523, A1
\bibitem[Joachimi \& Schneider (2008)]{20a} Joachimi B., Schneider P., 2008, A\&A, 488, 829
\bibitem[Joachimi \& Schneider (2009)]{20b} Joachimi B., Schneider P., 2009, A\&A, 507, 105
\bibitem[Joachimi \& Schneider (2010)]{20c} Joachimi B., Schneider P., 2010, A\&A, 517, A4
\bibitem[Joachimi et~al. (2010)]{15} Joachimi B., Mandelbaum R., Abdalla F., Bridle S., 2010, A\&A, 527, A26
\bibitem[Joudaki, Cooray \& Holz (2009)]{1i} Joudaki S., Cooray A., Holz D.E., 2009, PRD, 80, 023003,
\bibitem[King (2005)]{17} King L., 2005, A\&A, 441, 47
\bibitem[King \& Schneider (2002)]{6j} King L., Schneider P., 2002, A\&A, 396, 411 
\bibitem[King \& Schneider (2003)]{18b} King L., Schneider P., 2003, A\&A, 398, 23
\bibitem[Kirk, Bridle \& Schneider (2010)]{19} Kirk D., Bridle S., Schneider M., 2010, MNRAS, 408, 1502
\bibitem[Komatsu et~al. (2011)]{33} Komatsu E., et~al., 2011, ApJS, 192, 18
\bibitem[Krause \& Hirata (2011)]{8h} Krause E., Hirata C.M., 2011, MNRAS, 410, 2730
\bibitem[Linder \& Cahn (2007)]{2j} Linder E., Cahn R., 2007, Astropart. Phys. 28, 481 
\bibitem[LSST Science Collaborations and LSST Project (2009)]{29} LSST Science Collaborations and LSST Project, 2009, LSST Science Book, Version 2.0, arXiv:astro-ph/0912.0201, http://www.lsst.org/lsst/scibook/
\bibitem[Mandelbaum et~al. (2006)]{12} Mandelbaum R., Hirata C.M., Ishak M., Seljak U., Brinkmann J., 2006, MNRAS, 367, 611
\bibitem[Massey et~al. (2005)]{1j} Massey R., Refregier A., Bacon D., Ellis R., 2005, MNRAS, 359, 1277
\bibitem[Okumura T., Jing (2009)]{14b} Okumura T., Jing Y.P., 2009, ApJL, 694, L83
\bibitem[Pen et~al. (2003)]{1k} Pen U.-L., Lu T., Van Waerbeke L., Mellier Y., 2003, MNRAS, 346, 994
\bibitem[Refregier (2003)]{6k} Refregier A., 2003, Ann. Rev. A\&A, 41, 645
\bibitem[Rhodes, Refregier \& Groth (2001)]{1l} Rhodes J., Refregier A., Groth E.J., 2001, ApJL, 552, L85 
\bibitem[Schmidt (2008)]{2k} Schmidt F., 2008, PRD, 78, 043002
\bibitem[Schrabback et~al. (2010)]{1m} Schrabback T. et~al., 2010, A\&A, 516, A63
\bibitem[Scoccimarro \& Couchman (2001)]{30} Scoccimarro R., Couchman H., 2001, MNRAS, 325, 1312
\bibitem[Semboloni et~al. (2008)]{16} Semboloni E., Heymans C., Van Waerbeke L., Schneider P., 2008, MNRAS, 388, 991
\bibitem[Semboloni et~al. (2010)]{5} Semboloni E., Schrabback T., Van Waerbeke L., Vafaei S., Hartlap J., Hilbert S., 2010, MNRAS, 410, 143
\bibitem[Simpson et~al. (2011)]{31} Simpson F., James J., Heavens A., Heymans C., 2011, arXiv:astro-ph/1107.5169
\bibitem[Shi, Joachimi \& Schneider (2010)]{21} Shi X., Joachimi B., Schneider P., 2010, A\&A, 523, A60
\bibitem[Song (2005)]{2l} Song Y.S., 2005, PRD, 71, 024026
\bibitem[Takada \& Jain (2003)]{3a} Takada M., Jain B., 2003, MNRAS, 340, 580 
\bibitem[Takada \& Jain (2004)]{4} Takada B., Jain M., 2004, MNRAS, 348, 897 
\bibitem[Takada \& White (2004)]{6l} Takada M., White M., 2004, ApJL, 601, L1 
\bibitem[Thomas, Abdalla \& Weller (2009)]{2m} Thomas S., Abdalla F., Weller J., 2009, MNRAS, 395, 197
\bibitem[Toreno, Semboloni \& Schrabback (2010)]{2n} Toreno I., Semboloni E., Schrabback T., 2010, A\&A, 530, A68
\bibitem[Vafaei et~al. (2010)]{3b} Vafaei S., Lu T., Van Waerbeke L., Semboloni E., Heymans C., Pen U.-L., 2010, Astropart. Phys. 32, 340
\bibitem[Van Waerbeke \& Mellier (2003)]{6m} Van Waerbeke L., Mellier Y., 2003, ArXiv Astrophysics e-prints, astro-ph/0305089 
\bibitem[Van Waerbeke et~al. (2000)]{1n} Van Waerbeke L. et~al., 2000, A\&A, 358, 30
\bibitem[Van Waerbeke et~al. (2002)]{1o} Van Waerbeke L., Mellier Y., Pell R., Pen U.-L., McCracken H.J., Jain B., 2002, A\&A, 393, 369 
\bibitem[Zaldarriaga, Spergel \& Seljak (1997)]{1p} Zaldarriaga M., Spergel D.N., Seljak U., 1997, ApJ, 488, 1 
\bibitem[Zhang (2010a)]{22} Zhang P., 2010, ApJ, 720, 1090 
\bibitem[Zhang (2010b)]{23} Zhang P., 2010, MNRAS, 406, L95
\bibitem[Zhang et~al. (2007)]{2o} Zhang P.J., Liguori M., Bean R., Dodelson S., 2007, PRL, 99, 141302
\bibitem[Zhang, Pen \& Bernstein (2010)]{36} Zhang P., Pen U.-L., Bernstein G., 2010, MNRAS, 405, 359
\bibitem[Zhao et~al. (2006)]{2p} Zhao H., Bacon D.J., Taylor A.N., Horne K., 2006, MNRAS, 368, 171
\bibitem[Zhao et~al. (2009)]{2q} Zhao G., Pogosian L., Silvestri A., Zylberberg J., 2009, PRD, 79, 083513

\end{thebibliography}
\end{document}